\documentclass[useAMS,usenatbib,letterpaper]{mn2e}
\usepackage{txfonts}
\usepackage{graphicx}
\usepackage{natbib}
\bibpunct{(}{)}{;}{a}{}{,}

\newcommand{\eps}{\varepsilon}

\newcommand{\p}{{\rm p}}

\renewcommand{\mp}{m_\p}
\newcommand{\CR}{{\rm CR}}
\renewcommand{\th}{{\rm th}}

\newcommand{\bra}{\langle}
\newcommand{\ket}{\rangle}

\newcommand{\B}{{\mathcal B}}
\newcommand{\M}{{\mathcal M}}
\newcommand{\dd}{\mathrm{d}}

\newcommand{\eff}{\mathrm{eff}}
\newcommand{\vecbf}{\mathbfit}
\newcommand{\dps}{\displaystyle}
\newcommand{\vel}{\upsilon}
\newcommand{\A}{\tilde{A}}
\newcommand{\s}{\rmn{s}}
\newcommand{\GADGET}{{\sc gadget}}

\def\del#1{{}}

\title[Shocks in cosmological SPH simulations]{Detecting shock waves in
  cosmological smoothed particle hydrodynamics simulations}

\author[C.~Pfrommer, V.~Springel, T.~A.~En{\ss}lin, M.~Jubelgas]
{Christoph~Pfrommer,$^{1,\,2}$\thanks{e-mail: pfrommer@cita.utoronto.ca (CP);
    volker@mpa-garching.mpg.de (VS); ensslin@mpa-garching.mpg.de (TAE);
    jubelgas@mpa-garching.mpg.de (MJ)} Volker~Springel,$^1$\footnotemark[1]
  Torsten~A.~En{\ss}lin,$^1$\footnotemark[1] Martin~Jubelgas$^1$\footnotemark[1]
  \\
  $^1$Max-Planck-Institut f\"ur Astrophysik, Karl-Schwarzschild-Stra{\ss}e 1,
  Postfach 1317, 85741 Garching, Germany \\
  $^2$Canadian Institute for Theoretical Astrophysics, University of Toronto,
  60 St. George Street, Toronto, Ontario, M5S 3H8, Canada }

\begin{document}
\pagerange{\pageref{firstpage}--\pageref{lastpage}} \pubyear{2005}
\maketitle
\label{firstpage}

\begin{abstract}
  We develop a formalism for the identification and accurate estimation of the
  strength of structure formation shocks {\em during} cosmological smoothed
  particle hydrodynamics simulations. Shocks not only play a decisive role for
  the thermalization of gas in virialising structures but also for the
  acceleration of relativistic cosmic rays (CRs) through diffusive shock
  acceleration. Our formalism is applicable both to ordinary non-relativistic
  thermal gas, and to plasmas composed of CRs and thermal gas. To this end, we
  derive an analytical solution to the one-dimensional Riemann shock tube
  problem for a composite plasma of CRs and thermal gas. We apply our methods
  to study the properties of structure formation shocks in high-resolution
  hydrodynamic simulations of the Lambda cold dark matter ($\Lambda$CDM) model.
  We find that most of the energy is dissipated in weak internal shocks with
  Mach numbers $\M\sim 2$ which are predominantly central flow shocks or merger
  shock waves traversing halo centres. Collapsed cosmological structures are
  surrounded by external shocks with much higher Mach numbers up to $\M\sim
  1000$, but they play only a minor role in the energy balance of
  thermalization.  This is because of the higher pre-shock gas densities within
  non-linear structures, and the significant increase of the mean shock speed
  as the characteristic halo mass grows with cosmic time.  We show that after
  the epoch of cosmic reionisation the Mach number distribution is
  significantly modified by an efficient suppression of strong external shock
  waves due to the associated increase of the sound speed of the diffuse
  gas. Invoking a model for CR acceleration in shock waves, we find that the
  average strength of shock waves responsible for CR energy injection is higher
  than that for shocks that dominate the thermalization of the gas. This implies
  that the dynamical importance of shock-injected CRs is comparatively large in
  the low-density, peripheral halo regions, but is less important for the weaker
  flow shocks occurring in central high-density regions of haloes. When combined
  with radiative dissipation and star formation, our formalism can also be used
  to study CR injection by supernova shocks, or to construct models for
  shock-induced star formation in the interstellar medium.
\end{abstract}
\begin{keywords}Shock waves -- intergalactic medium -- galaxies:
clusters: general -- cosmology: large-scale structure of universe -- cosmic
rays -- methods: numerical
\end{keywords}

\section{Introduction}
\label{sec:intro}

\subsection{Structure formation shock waves}

Cosmological shock waves form abundantly in the course of structure formation,
both due to infalling pristine cosmic plasma which accretes onto filaments,
sheets and haloes, as well as due to supersonic flows associated with merging
substructures \citep{1998ApJ...502..518Q, 2000ApJ...542..608M,
  2003ApJ...593..599R, 2003ApJ...583..695G, 2005astro.ph..4485P}.
Additionally, shock waves occur due to non-gravitational physics in the
interstellar and intracluster media, e.g.~as a result of supernova explosions.
Structure formation shock waves propagate through the cosmic tenuous plasma,
which is compressed at the transition layer of the shock while a part of the
kinetic energy of the incoming plasma is dissipated into internal energy of
the post-shock gas.  Because of the large collisional mean free path, the
energy transfer proceeds through collective electromagnetic viscosity which is
provided by ubiquitous magnetic irregularities
\citep{1974ARA&A..12...71W,1985GMS....34....1K}.

Cosmologically, shocks are important in several respects. (1) Shock waves
dissipate gravitational energy associated with hierarchical clustering into
thermal energy of the gas contained in dark matter haloes, thus supplying the
intrahalo medium with entropy and thermal pressure support. Radiative
cooling is then required to compress the gas further to densities that will
allow star formation. (2) Shocks also occur around moderately overdense
filaments, which leads to a heating of the intragalactic medium.  Sheets and
filaments are predicted to host a warm-hot intergalactic medium with
temperatures in the range $10^5\,\mbox{K}<T<10^7\,\mbox{K}$ whose evolution is
primarily driven by shock heating from gravitational perturbations breaking on
mildly nonlinear, non-equilibrium structures \citep{1998ApJ...509...56H,
  1999ApJ...514....1C, 2001ApJ...552..473D, 2004ApJ...611..642F,
  2005ApJ...620...21K}.  Thus, the shock-dissipated energy traces the large
scale structure and contains information about its dynamical history.  (3)
Besides thermalization, collisionless shocks are also able to accelerate ions
of the high-energy tail of the Maxwellian through diffusive shock acceleration
(DSA) \citep[for reviews see][]{1983RPPh...46..973D, 1987PhR...154....1B,
  2001RPPh...64..429M}. These energetic ions are reflected at magnetic
irregularities through magnetic resonances between the gyro-motion and waves
in the magnetised plasma and are able to gain energy in moving back and forth
through the shock front. This acceleration process typically yields a cosmic
ray (CR) population with a power-law distribution of the particle momenta.
Nonlinear studies of DSA have shown that a considerable part of the kinetic
energy flux passing through shocks can be channelled into non-thermal
populations, up to about one-half of the initial kinetic energy of the shock
\citep{1995NuPhS..39A.171B, 1996ApJ...473.1029E, 1998PhRvE..58.4911M,
  1999ApJ...511L..53M, 2002ApJ...579..337K}.  Note that CRs have sufficient
momentum not to resonate with the electromagnetic turbulence in the shock
front itself. They hence experience the shock as a discontinuity, i.e.~the CR
population is adiabatically compressed by the shock
\citep[e.g.,][]{1983RPPh...46..973D}.

Indeed, CR electrons have been observed in the intra-cluster medium (ICM) of
galaxy clusters through their diffuse synchrotron emission
\citep{1989Natur.341..720K, 1993ApJ...406..399G, 1997A&A...321...55D}. In
addition to these extended radio haloes which show a similar morphology
compared to the thermal X-ray emission, there have been extended radio relics
observed in the cluster periphery \citep[e.g.,][]{1997MNRAS.290..577R} which
might well coincide with merger shock waves as proposed by
\citet{1998AA...332..395E}.  Some clusters have also been reported to exhibit
an excess of hard X-ray emission compared to the expected thermal
bremsstrahlung of the hot ICM, most probably produced by inverse Compton
up-scattering of cosmic microwave background photons by relativistic electrons
\citep{1999ApJ...513L..21F, 2005MNRAS.360..133S}. It has been proposed that a
fraction of the diffuse cosmological $\gamma$-ray background radiation
originates from the same processes \citep{2000Natur.405..156L,
  2002MNRAS.337..199M, 2003ApJ...588..155R, 2003ApJ...594..709B,
  2005ApJ...618..675K}.

To date, there are two different scenarios explaining these non-thermal
emission processes. (1) Reacceleration processes of `mildly' relativistic
electrons ($\gamma\simeq 100-300$) being injected over cosmological timescales
into the ICM by sources like radio galaxies, supernova remnants, merger
shocks, or galactic winds, which all can provide an efficient supply of
highly-energetic CR electrons.  Owing to their long lifetimes of a few times
$10^9$ years these `mildly' relativistic electrons can accumulate within the
ICM \citep{2002mpgc.book....1S}, until they experience continuous in-situ
acceleration either via shock acceleration or resonant pitch angle scattering
on turbulent Alfv\'en waves \citep{1977ApJ...212....1J, 1987A&A...182...21S,
  2001MNRAS.320..365B, 2002ApJ...577..658O, 2004MNRAS.350.1174B}. (2) In the
ICM, the CR protons have lifetimes of the order of the Hubble time
\citep{1996SSRv...75..279V}, which is long enough to diffuse away from the
production site and to maintain a space-filling distribution over the cluster
volume.  These CR protons can interact hadronically with the thermal ambient
gas producing secondary electrons, neutrinos, and $\gamma$-rays in inelastic
collisions throughout the cluster volume, generating radio haloes through
synchrotron emission \citep{1980ApJ...239L..93D, 1982AJ.....87.1266V,
  1999APh....12..169B, 2000A&A...362..151D, 2003A&A...407L..73P,
  2004A&A...413...17P, 2004MNRAS.352...76P}. Cosmological simulations support
the possibility of a hadronic origin of cluster radio haloes
\citep{2001ApJ...562..233M}.

\subsection{Hydrodynamical simulations}

Hydrodynamical solvers of cosmological codes are generally classified into two
main categories: (1) Lagrangian methods like smoothed particle hydrodynamics
(SPH) which discretise the mass of the fluid, and (2) Eulerian codes, which
discretise the fluid volume. SPH methods were first proposed by
\citet{1977MNRAS.181..375G} and \citet{1977AJ.....82.1013L} and approximate
continuous fluid quantities by means of kernel interpolation over a set of
tracer particles. Over the years, SPH techniques have been steadily improved
and found widespread applications in cosmological problems
\citep{1988MNRAS.235..911E, 1989ApJS...70..419H, 1993MNRAS.265..271N,
2002MNRAS.333..649S}.

In contrast, Eulerian methods discretise space and represent continuous fields
on a mesh.  Originally, Eulerian codes employed a mesh which is fixed in space
\citep{1993ApJ...417..404C, 1995clun.conf..209Y} or adaptively moving
\citep{1998ApJS..115...19P}, while more recently, adaptive mesh refinement
(AMR) algorithms have been developed for cosmological applications
\citep{1989JCP...82..64B, 1997ASPC..123..363B, 1999numa.conf...19N,
2002Sci...295...93A, 2002ApJ...571..563K, 2002PhRvD..66d3002R}, which can adapt
to regions of interest in a flexible way.

Grid-based techniques offer superior capabilities for capturing hydrodynamical
shocks. In some algorithms, this can be done even without the aid of artificial
viscosity, thanks to the use of Riemann solvers at the cell-level, so that a
very low residual numerical viscosity is achieved. However, codes employing
static meshes still lack the resolution and flexibility necessary to tackle
structure formation problems in a hierarchically clustering universe, which is
characterised by a very large dynamic range and a hierarchy of substructure at
all stages of the evolution.  For example, techniques based on a fixed mesh are
seriously limited when one tries to study the formation of individual galaxies
in a cosmological volume, simply because the internal galactic structure such
as disk and bulge components can then in general not be sufficiently well
resolved.  A new generation of AMR codes which begin to be applied in cosmology
may in principle resolve this problem. However, a number of grid-based problems
remain even here, for example the dynamics is not Galilean-invariant, and there
can be spurious advection and mixing errors, especially for large bulk
velocities across the mesh.

These problems can be avoided in SPH, which thanks to its Lagrangian nature
and its accurate treatment of self-gravity is particularly well suited for
structure formation problems. SPH adaptively and automatically increases the
resolution in dense regions such as galactic haloes or centres of galaxy
clusters, which are the regions of primary interest in cosmology.  One
drawback of SPH is the dependence on the artificial viscosity which has to
deliver the necessary entropy injection in shocks. While the parametrization
of the artificial viscosity can be motivated in analogy with the Riemann
problem \citep{1997JCoPh.136..298M}, the shocks themselves are broadened over
the SPH smoothing scale and not resolved as discontinuities, but post-shock
quantities are calculated very accurately. However, to date it has not been
possible to identify and measure the shock strengths instantaneously with an
SPH simulation.

Being interested in dynamical implications of CRs on structure formation and
galaxy evolution, one faces not only the problem of the interplay of gravity
and hydrodynamics of a plasma composed of CRs and thermal particles but in
addition radiative processes such as cooling and supernova feedback. To date,
AMR codes have not yet matured to the point that they can address all these
requirements throughout a cosmological volume, although there are recent
efforts along these lines \citep[e.g.][]{2005ApJ...620...44K,
2005APh....24...75J}.  It would therefore be ideal if SPH codes for structure
formation could acquire the ability to detect shocks reliably {\em during}
simulations. Previous work on shock detection in SPH simulations
\citep{2003ApJ...585..128K} was restricted to a posteriori analysis of two
subsequent simulation time-slices, which can then be used to approximately
detect a certain range of shocks as entropy jumps.

\subsection{Motivation and structure}

This article seeks to close this gap in order to allow studies of the following
questions. (1) The cosmic evolution of shock strengths provides rich
information about the thermal history of the baryonic component of the
Universe: where and when is the gas heated to its present temperatures, and
which shocks are mainly responsible for it?  Does the missing baryonic
component in the present-day universe reside in a warm-hot intergalactic
medium?  (2) CRs are accelerated at structure formation shocks through
diffusive shock acceleration: what are the cosmological implications of such a
CR component?  (3) Shock waves are modified by nonlinear back-reaction of the
accelerated CRs and their spatial diffusion into the pre-shock regime: does
this change the cosmic thermal history or give rise to other effects?  (4)
Simulating realistic CR profiles within galaxy clusters can provide detailed
predictions for the expected radio synchrotron and $\gamma$-ray emission. What
are the observational signatures of this radiation that is predicted to be
observed with the upcoming new generation of $\gamma$-ray instruments (imaging
atmospheric \v{C}erenkov telescopes and the GLAST\footnote{Gamma-ray Large Area
Space Telescope, http://glast.gsfc.nasa.gov/} satellite) and radio
telescopes (LOFAR\footnote{LOw Frequency ARray,
http://www.lofar.org/} and extended Very Large Array)?

The purpose of this paper is to study the properties of structure formation
shock waves in cosmological simulations, allowing us to explore their role for
the thermalization of the pristine plasma, as well as for the acceleration of
relativistic CRs through DSA.  In particular, we develop a framework for
quantifying the importance of CRs during cosmological structure formation,
including an accounting of the effects of adiabatic compressions and
rarefactions of CR populations, as well as of numerous non-adiabatic
processes. Besides CR injection by structure formation shocks, the latter
include CR shock injection of supernova remnants, in-situ re-acceleration of
CRs, spatial diffusion of CRs, CR energy losses due to Coulomb interactions,
Bremsstrahlung, and hadronic interactions with the background gas, and the
associated $\gamma$-ray and radio emission due to subsequent pion decay. A
full description of these CR processes and their formulation for cosmological
applications is described in \citet{2005...Ensslin}, while the numerical
implementation within the SPH formalism is given by \citet{2005...Jubelgas}.
In this work we provide a crucial input for this modelling: a formalism for
identifying and accurately estimating the strength of structure formation
shocks on-the-fly during cosmological SPH simulations.

The paper is structured as follows. The basic cosmic ray variables are
introduced in Section~\ref{sec:bascis}. The formalism for identifying and
measuring the Mach number of shock waves instantaneously within an SPH
simulation is described in Section~\ref{sec:formalism} for a purely thermal gas
as well as for a composite plasma of CRs and thermal gas. The numerical
implementation of the algorithm is discussed in Section~\ref{sec:numimpl}. In
Section~\ref{sec:shocktube}, we compare shock tube simulations to analytic
solutions of the Riemann problem which are presented in
Appendices~\ref{sec:Riemann} and \ref{sec:Riemann+CRs}. Finally, in
Section~\ref{sec:simulations}, we perform cosmological non-radiative simulations
to study CR energy injection at shocks, and the influence of reionisation on
the Mach number distribution.  A summary in Section~\ref{sec:summary} concludes
the paper.

\section{Basic cosmic ray variables}
\label{sec:bascis}

Since we only consider CR protons\footnote{$\alpha$-particles carry a
  significant fraction of the total CR energy.  Nevertheless, the assumption of
  considering only CR protons is a reasonable approximation, since the energy
  density of $\alpha$-particles can be absorbed into the proton spectrum.  A
  GeV energy $\alpha$-particle can be approximated as an ensemble of four
  individual nucleons travelling together due to the relatively weak MeV nuclear
  binding energies compared to the kinetic energy of relativistic protons.},
which are at least in our Galaxy the dominant CR species, it is convenient to
introduce the dimensionless momentum $p = P_\p/(\mp\,c_\rmn{light})$.  CR
electrons with $\gamma < 100$ experience efficient Coulomb losses such that
their energy density is significantly diminished compared to the CR energy
density \citep{2002mpgc.book....1S}.  The differential particle momentum
spectrum per volume element is assumed to be a single power-law above the
minimum momentum $q$:
\begin{equation}
\label{eq:spec1}
f(p) = \frac{\dd N}{\dd p\,\dd V} = C \, p^{-\alpha}\,
\theta(p- q) .
\end{equation}
$\theta(x)$ denotes the Heaviside step function. Note that we use an effective
one-dimensional distribution function $f(p)\equiv 4\pi p^2 f^{(3)}(p)$.  The
CR population can hydrodynamically be described by an isotropic pressure
component as long as the CRs are coupled to the thermal gas by small scale
chaotic magnetic fields.  The differential CR spectrum can vary spatially and
temporally (although for brevity we suppress this in our notation) through the
spatial dependence of the normalisation $C=C(\vecbf{r},t)$ and the cutoff $q =
q(\vecbf{r},t)$.

Adiabatic compression or expansion leaves the phase-space density of the CR
population unchanged, leading to a momentum shift according to $p \rightarrow
p' = (\rho/\rho_0)^{1/3}\, p$ for a change in gas density from $\rho_0$ to
$\rho$.  Since this is fully reversible, it is useful to introduce the
invariant cutoff and normalisation $q_0$ and $C_0$ which describe the CR
population via equation  (\ref{eq:spec1}) if the inter-stellar medium (ISM) or ICM
is adiabatically compressed or expanded to the reference density $\rho_0$. The
actual parameters are then given by
\begin{equation}
\label{eq:adiabatic}
q(\rho) = \left(\frac{\rho}{\rho_0} \right)^{{1}/{3}}\, q_0\;\;
\mbox{and}\;\; C(\rho) = \left(\frac{\rho}{\rho_0}
\right)^{({\alpha+2})/{3}}\, C_{0}.
\end{equation}
These adiabatically invariant variables are a suitable choice to be used in a
Lagrangian description of the CR population.

The CR number density is
\begin{equation}
\label{eq:ncr}
n_{\CR} = \int_0^\infty  \dd p\, f(p) =
\frac{C\, q^{1-\alpha}}{\alpha-1} ,
\end{equation}
provided, that $\alpha >1$.
The kinetic energy density of the CR population is
\begin{eqnarray}
\label{eq:epscr}
\eps_\CR &=& \int_0^\infty \dd p\, f(p) \,T_{\rm
p}(p)=\frac{C\, \mp\,c_\rmn{light}^2}{\alpha-1} \, \times
\nonumber \\
&& \left[\frac{1}{2}
\, \B_{\frac{1}{1+q^2}} \left(
\frac{\alpha-2}{2},\frac{3-\alpha}{2}\right) + q^{1-\alpha}
\left(\sqrt{1+q^2}-1 \right) \right] \,,
\end{eqnarray}
where $T_\p(p) = (\sqrt{1+p^2} -1)\, \mp\,c_\rmn{light}^2$ is the kinetic
energy of a proton with momentum $p$, and $\B_x(a,b)$ denotes the incomplete
Beta-function which is defined by $\B_x(a,b)\equiv\int_0^x t^{a-1}
(1-t)^{b-1} \dd t$. The integral of equation~(\ref{eq:epscr}) is well-defined if we
assume $\alpha>2$.  The CR pressure is
\begin{eqnarray}
\label{eq:Pcr}
P_\CR &=& \frac{\mp c_\rmn{light}^2}{3}\!\int_0^\infty \! \dd p\, f(p)
\,\beta\,p  \nonumber\\
&=&\frac{C\,\mp c_\rmn{light}^2}{6} \, 
\B_{\frac{1}{1+q^2}} \left( \frac{\alpha-2}{2},\frac{3-\alpha}{2}
\right),
\end{eqnarray}
where $\beta \equiv \vel/c_\rmn{light} = p/\sqrt{1+p^2}$ is the dimensionless
velocity of the CR particle. Note that for $2<\alpha<3$ the kinetic energy
density and pressure of the CR populations are well defined for the limit
$q\rightarrow 0$, although the total CR number density diverges.

The adiabatic exponent of the CR population is defined by 
\begin{equation}
\label{eq:gammaCR}
\gamma_\CR \equiv \left.\frac{\dd \log P_\CR}{\dd \log \rho}\right|_S,
\end{equation}
while the derivative has to be taken at constant entropy $S$. Using
equations~(\ref{eq:adiabatic}) and (\ref{eq:Pcr}), we obtain for the CR adiabatic
exponent
\begin{eqnarray}
\label{eq:gammaCR2}
\gamma_\CR &=& \frac{\rho}{P_\CR}
\left(\frac{\upartial P_\CR}{\upartial C}\frac{\upartial C}{\upartial \rho}
+ \frac{\upartial P_\CR}{\upartial q}\frac{\upartial q}{\upartial \rho}\right) \nonumber\\
&=& \frac{\alpha + 2}{3} - \frac{2}{3}\,
q^{2-\alpha}\, \beta(q)
\left[\B_{\frac{1}{1+q^2}} 
 \left( \frac{\alpha-2}{2},\frac{3-\alpha}{2}\right) \right]^{-1}.
\end{eqnarray}
Note that in contrast to the usual adiabatic exponent, the CR adiabatic
exponent is time dependent due to its dependence on the lower cutoff of the CR
population, $q$. The ultra-relativistic limit ($q\to\infty$) of the
adiabatic exponent, where $\gamma_\CR \to 4/3$, can easily be obtained by using
the integral representation of the incomplete Beta-function and applying a
Taylor expansion to the integrand. In the non-relativistic limit ($q\ll 1$ and
$\alpha>3$), the adiabatic exponent approaches $\gamma_\CR \to 5/3$. This can
be seen by evaluating the CR pressure in this limit, $P_\CR = \frac{m_\p
c_\rmn{light}^2}{3\, (\alpha-3)}\,C\,q^{3-\alpha}$ and applying the definition of
$\gamma_\CR$ in equation~(\ref{eq:gammaCR}).  Considering a composite of thermal
and CR gas, it is appropriate to define an effective adiabatic index by
\begin{equation}
\label{eq:gammaeff}
\gamma_\rmn{eff} \equiv \left.\frac{\dd \log (P_\rmn{th} + P_\CR)}
{\dd \log \rho}\right|_S = 
\frac{\gamma_\rmn{th}\, P_\rmn{th} + \gamma_\CR\, P_\CR}
{P_\rmn{th} + P_\CR}.
\end{equation}

\section{Mach numbers within the SPH formalism}
\label{sec:formalism}

The shock surface separates two regions: the {\em upstream regime} (pre-shock
regime) defines the region in front of the shock whereas the {\em downstream
  regime} (post-shock regime) defines the wake of the shock wave.  The shock
front itself is the region in which the mean plasma velocity changes rapidly on
small scales given by plasma physical processes.  All calculations in this
section are done in the rest frame of the shock which we assume to be
non-relativistic. This assumption is justified in the case of cosmological
structure formation shock waves for which typical shock velocities are of the
order of $10^3 \mbox{ km s}^{-1}$.

Particles are impinging on the shock surface at a rate per unit shock surface,
$j$, while conserving their mass:
\begin{equation}
\label{eq:mass_conservation}
\rho_1 \vel_1 = \rho_2 \vel_2 = j.
\end{equation}
Here $\vel_1$ and $\vel_2$ indicate the plasma velocities (relative to the shock's
rest frame) in the upstream and downstream regime of the shock, respectively.
The mass densities in the respective shock regime are denoted by $\rho_1$ and
$\rho_2$.  Momentum conservation implies
\begin{equation}
\label{eq:momentum_conservation}
P_1 + \rho_1 \vel_1^2 = P_2 + \rho_2 \vel_2^2,
\end{equation}
where $P_i$ denotes the pressure  in the respective regime $i \in \{1,2\}$.
The energy conservation law at the shock surface reads
\begin{equation}
\label{eq:energy_conservation}
(\eps_1 + P_1)\, \rho_1^{-1} + \frac{\vel_1^2}{2} = 
(\eps_2 + P_2)\, \rho_2^{-1} + \frac{\vel_2^2}{2}. 
\end{equation}
$\eps_i$ denotes the internal energy density in the regime $i \in \{1,2\}$.
Combining solely these three equations without using any additional information
about the equation of state, we arrive at the following system of two
equations:
\begin{eqnarray}
\label{eq:j_squared}
j^2 &=& 
\rho_1^2 \M_1^2 c_1^2 = \frac{(P_2 - P_1)\, \rho_1 \rho_2}{\rho_2 - \rho_1} \\
\label{eq:rho_general}
\frac{\rho_2}{\rho_1} &=&
\frac{2 \eps_2 + P_1 + P_2}{2 \eps_1 + P_1 + P_2}.
\end{eqnarray}
Here we introduced the Mach number in the upstream regime, $\M_1 = \vel_1/c_1$,
which is the plasma velocity in units of the local sound speed $c_1 =
\sqrt{\gamma P_1/\rho_1}$.\footnote{Note, that the symbol $c$ (sometimes with
  subscript) denotes the sound velocity. }

\subsection{Polytropic gas}

Non-radiative polytropic gas in the regime $i \in \{1,2\}$ is
characterised by its particular equation of state,
\begin{equation}
\label{eq:eos}
\eps_i = \frac{1}{\gamma-1} P_i
\quad\mbox{or equivalently}\quad
P_i = P_0 \left(\frac{\rho_i}{\rho_0}\right)^\gamma,
\end{equation}
where $\gamma$ denotes the adiabatic index. This allows us to derive the
well-known Rankine-Hugoniot conditions which relate quantities from the
upstream to the downstream regime solely as a function of $\M_1$:
\begin{eqnarray}
\label{eq:rho_jump}
\frac{\rho_2}{\rho_1} &=&
\frac{(\gamma+1) \M_1^2}{(\gamma-1) \M_1^2 + 2}, \\
\label{eq:P_jump}
\frac{P_2}{P_1} &=&
\frac{2\gamma \M_1^2 - (\gamma-1)}{\gamma+1}, \\
\label{eq:T_jump}
\frac{T_2}{T_1} &=&
\frac{\left[2\gamma\M_1^2 - (\gamma-1)\right]
      \left[(\gamma-1)\M_1^2 + 2\right]}
{(\gamma+1)^2 \M_1^2}.
\end{eqnarray}

In cosmological simulations using a Lagrangian description of
hydrodynamics such as SPH, it is infeasible to identify the rest frame
of each shock and thus $\M_1$ unambiguously, especially in the
presence of multiple oblique structure formation shocks. As an
approximative solution, we rather propose the following procedure,
which takes advantage of the entropy-conserving formulation of SPH
\citep{2002MNRAS.333..649S}. For one particle, the instantaneous
injection rate of the entropic function due to shocks is computed,
i.e. $\dd A /\dd t$, where $A$ denotes the entropic function $A(s)$
defined by
\begin{equation}
P = A(s)\rho^\gamma,
\end{equation}
and $s$ gives the specific entropy.  Suppose further that the shock is
broadened to a scale of order the SPH smoothing length $f_h h$, where $f_h \sim
2$ denotes a factor which has to be calibrated against shock-tubes. We can
roughly estimate the time it takes the particle to pass through the broadened
shock front as $\Delta t = f_h h/\vel$, where one may approximate $\vel$ with
the pre-shock velocity $\vel_1$. Assuming that the present particle temperature
is a good approximation for the pre-shock temperature, we can also replace
$\vel_1$ with $\M_1 c_1$.\footnote{Extensions of this approach that apply
to particles within the SPH broadened shock surface will be described in
Section~\ref{sec:numimpl_th}.}  

Based on these assumptions and using $\Delta A_1\simeq\Delta t\, \dd A_1/\dd
t$, one can estimate the jump of the entropic function the particle will
receive while passing through the shock:
\begin{eqnarray}
\label{eq:A2/A1_1}
\frac{A_2}{A_1}  &=& \frac{A_1 + \Delta A_1}{A_1} = 1 + \frac{f_h h}{\M_1 c_1 A_1}
\frac{\dd A_1}{\dd t}, \\
\label{eq:A2/A1_2}
\frac{A_2}{A_1}  &=& 
\frac{P_2}{P_1} \left(\frac{\rho_1}{\rho_2}\right)^{\gamma}
= f_A(\M_1),
\end{eqnarray}
where
\begin{equation}
f_A(\M_1) \equiv \frac{2\gamma \M_1^2 - (\gamma-1)}{\gamma+1}
\left[\frac{(\gamma-1) \M_1^2 + 2}{(\gamma+1) \M_1^2}\right]^\gamma,
\end{equation}
using equations~(\ref{eq:rho_jump}) and (\ref{eq:P_jump}). Combining
equations~(\ref{eq:A2/A1_1}) and (\ref{eq:A2/A1_2}), we arrive at the final
equation which is a function of Mach number only:
\begin{equation}
\left[f_A(\M_1) - 1 \right] \M_1 = 
\frac{f_h h}{c_1 A_1} \frac{\dd A_1}{\dd t}.
\end{equation}
The right-hand side can be estimated individually for each particle, and the
left-hand side depends only on $\M_1$. Determining the root of the equation hence
allows one to estimate a Mach number for each particle.

\subsection{Composite of cosmic rays and thermal gas}
\label{sec:SPHshock}

In the presence of a gas composed of cosmic rays and thermal components,
equations~(\ref{eq:mass_conservation}) to (\ref{eq:rho_general}) are still
applicable if one identifies the energy density $\eps_i$ and the pressure $P_i$
with the sum of the individual components in the regime  $i \in \{1,2\}$, 
\begin{eqnarray}
  \label{eq:composite}
  \eps_i &=& \eps_{\rmn{CR}i} + \eps_{\rmn{th}i}, \\
  P_i &=& P_{\rmn{CR}i} + P_{\rmn{th}i}.  
\end{eqnarray}
The sound speed of such a composite gas is $c_i = \sqrt{\gamma_{\rmn{eff},i}
  P_i/\rho_i}$, where $\gamma_{\rmn{eff},i}$ is given by equation~(\ref{eq:gammaeff}).
Note that in contrast to the single-component fluid, for the general case there is no
equivalent to the equation of state (equation~\ref{eq:eos}) in terms of the total
energy density $\eps_i$, because of the additivity of both pressure and energy
density. For later convenience, we introduce the shock compression ratio $x_\s$
and the thermal pressure ratio $y_\s$,
\begin{equation}
  \label{eq:x_y}
  x_\s = \frac{\rho_2}{\rho_1}
  \quad\mbox{and}\quad
  y_\s = \frac{P_\rmn{th2}}{P_\rmn{th1}}.
\end{equation}

While taking the equation of state (equation~\ref{eq:eos}) for the thermal gas
component, we assume adiabatic compression of the CRs at the shock\footnote{Due
  to their much larger gyro-radii and high velocities, CR protons should not
  participate in the plasma processes of collisionless shock waves.},
\begin{equation}
  \label{eq:CR_shock_cond}
  P_\rmn{CR2} = P_\rmn{CR1} x_\s^{\gamma_\rmn{CR}}
  \quad\mbox{and}\quad
  \eps_\rmn{CR2} = \eps_\rmn{CR1} x_\s^{\gamma_\rmn{CR}}.
\end{equation}
Here we assume a constant CR spectral index over the shock which holds only
approximately owing to the weak dependence of the CR lower momentum cutoff $q$ on
the density (equation~\ref{eq:adiabatic}). 

For the composite of thermal and CR gas, it is convenient to define the
effective entropic function $A_\rmn{eff}$ and its time derivative,
\begin{eqnarray}
  \label{eq:Aeff}
  A_\rmn{eff} &=& 
  \left(P_\rmn{th} + P_\rmn{CR}\right) \rho^{-\gamma_\rmn{eff}}, \\
  \label{eq:dAeff_dt}
  \frac{\dd A_\rmn{eff}}{\dd t} &=&
  \frac{\dd A_\rmn{th}}{\dd t} \rho^{\gamma_\rmn{th} - \gamma_\rmn{eff}}. 
\end{eqnarray}
The expression for the time derivative of the effective adiabatic function uses
the approximation of adiabatic compression of the CRs at the shock.  Using the
same assumptions like in the non-radiative case, we estimate the jump of the
entropic function for the particle on passing through the shock made of
composite gas:
\begin{equation}
\label{eq:Aeff_jump}
\frac{A_\rmn{eff,2}^{}}{A_\rmn{eff,1}^{}}  = 
\frac{\left(P_\rmn{CR2}^{} + P_\rmn{th2}^{}\right)\rho_2^{-\gamma_\rmn{eff,2}}}
     {\left(P_\rmn{CR1}^{} + P_\rmn{th1}^{}\right)\rho_1^{-\gamma_\rmn{eff,1}}} =
1 + \frac{f_h h}{\M_1 c_1 A_\rmn{eff,1}} \frac{\dd A_\rmn{eff,1}}{\dd t} .
\end{equation}

Combining equations (\ref{eq:j_squared}), (\ref{eq:rho_general}),
(\ref{eq:CR_shock_cond}), and (\ref{eq:Aeff_jump}), we arrive at the following
system of equations,
\begin{eqnarray}
  \label{eq:sys1}
  f_1(x_\s,y_\s) &=& 
  x_\s \big[P_2(x_\s,y_\s) - P_1\big] 
  \nonumber\\
  &\times&
  \left[P_2(x_\s,y_\s)(x_\s \rho_1)^{-\gamma_\rmn{\eff,2}(x_\s,y_\s)} -
    P_1^{} \rho_1^{-\gamma_\rmn{\eff,1}} \right]^2 
  \nonumber\\
  &-& 
  P_1^2 (x_\s-1)\, \rho_1^{1 - 2 \gamma_\rmn{\eff,1}}
  \left(\frac{f_h h}{A_\rmn{eff,1}} \frac{\dd A_\rmn{eff,1}}{\dd t}\right)^2 
  = 0, \\
  \label{eq:sys2}
  f_2(x_\s,y_\s) &=&
  2 \eps_2(x_\s,y_\s) + P_1 + P_2(x_\s,y_\s) 
  \nonumber\\
  &-& x_\s\,[2 \eps_1 + P_1 + P_2(x_\s,y_\s)] = 0.
\end{eqnarray}
The effective adiabatic index in the post-shock regime is given by
\begin{equation}
\label{eq:gamma2}
\gamma_\rmn{eff,2}(x_\s,y_\s) = 
\frac{\gamma_\rmn{CR} P_\rmn{CR2}(x_\s) + \gamma_\rmn{th} y_\s P_\rmn{th1}}
{P_2(x_\s,y_\s)}.
\end{equation}

Given all the quantities in the pre-shock regime, we can solve for the roots
$x_\s$ and $y_\s$ of this system of two non-linear equations.  This system of
equations turns out to be nearly degenerate for plausible values of pre-shock
quantities such that it might be convenient to apply the following coordinate
transformation:
\begin{equation}
\label{eq:x_z}
(x_\s,y_\s) \to (x_\s,z_\s) 
\quad \mbox{with} \quad
z_\s = \frac{y_\s - x_\s}{4}.
\end{equation}
The Mach number $\M_1$ and the jump of internal specific energies can then be
obtained by
\begin{eqnarray}
\label{eq:M1}
\M_1 &=& \sqrt{\frac{(P_2 - P_1) x_\s}{\rho_1 c_1^2 (x_\s - 1)}}
\quad\mbox{and} \\
\frac{u_2}{u_1} &=& \frac{y_\s}{x_\s}
\quad\mbox{where\quad} 
u = \frac{P_\th}{(\gamma_\th - 1)\, \rho}.
\end{eqnarray}

\section{Numerical implementation}
\label{sec:numimpl}

\subsection{Polytropic gas}
\label{sec:numimpl_th}
Applying the algorithm of inferring the shock strength within the SPH formalism
in a straightforward manner will lead to systematically underestimated values
of the Mach number for SPH particles which are located {\em within the SPH
broadened shock surface}: the proposed algorithm of Section~\ref{sec:formalism}
assumes that the present particle quantities such as entropy, sound velocity,
and smoothing length are good representations of the hydrodynamical state in
the pre-shock regime, which is not longer the case for particles within the SPH
broadened shock surface. To overcome this problem, we define a decay time
interval $\Delta t_\rmn{dec} = \mbox{min}[f_h h / (\M_1 c), \Delta
t_\rmn{max}]$, during which the Mach number is set to the maximum value that is
estimated during the transition from the pre-shock regime to the shock surface.
At this maximum, the corresponding particle quantities are good approximations
of the hydrodynamical values in the pre-shock regime. We thus have a finite
temporal resolution for detecting shocks, which is of order the transit time
through the broadened shock front. Note that $\Delta t_{\rm max}$ is just
introduced as a safeguard against too long decay times for very weak shocks. In
the case of cosmological simulations, which are conveniently carried out in a
computational domain that is comoving with the cosmological expansion, we
redefine the decay time $\Delta (\ln a)_\rmn{dec} = \mbox{min}[H(a) f_h h /
(\M_1 c), \Delta (\ln a)_\rmn{max}]$ where $a$ denotes the cosmic scale factor
and $H(a) = \dot{a}/a$ is the Hubble function. An appropriate choice for the
safeguard parameter is $\Delta (\ln a)_\rmn{max} = 0.0025$.

Secondly, there is no universal value $f_h$ which measures the SPH shock
broadening accurately irrespective of the Mach number of the shock, especially
in the regime of strong shocks.  We therefore use the original algorithm (with
$f_h = 2$) only for estimated Mach numbers with $\M_\rmn{est} < 3$, while for
stronger shocks, we apply an empirically determined formula (calibrated against
shock-tubes) which corrects for the additional broadening of strong shocks and
smoothly joins into the weak shock regime:
\begin{equation}
  \label{eq:calibration}
  \M_\rmn{cal} = \left(a_\M \M_\rmn{est}^{b_\M} + c_\M \exp^{-\M_\rmn{est} / 3}\right) 
  \M_\rmn{est},
\end{equation}
where $a_\M = 0.09$, $b_\M = 1.34$, and $c_\M = 1.66$. These numbers may depend
on the viscosity scheme of the SPH implementation.

\subsection{Composite of cosmic rays and thermal gas}

Our formalism of inferring the jump conditions for a composite of cosmic rays
and thermal gas yields the density jump, $x_\s = \rho_2 / \rho_1$, and the
thermal pressure jump at the shock, $y_\s = P_\rmn{th2} / P_\rmn{th1}$
(Section~\ref{sec:SPHshock}). As described in the previous section
(Section~\ref{sec:numimpl_th}), the values for the estimated jump conditions are
systematically underestimated in the regime of strong shocks ($\M_1 \gtrsim 5$)
implying an additional broadening of the shock surface. Thus, we proceed the
same way as above: using the value of the density jump $x_\s$, we derive the Mach number
of the shock through equation~(\ref{eq:M1}) and recalibrate it for strong shocks.
In addition, we use the decay time $\Delta t_\rmn{dec}$ as before in the
thermal case to obtain reliable Mach number estimates.  The post-shock density
is then obtained by multiplying the stored pre-shock density with the
density jump $x_\s$.

In the case of a thermal pressure jump at the shock $y_\s$, we decided not to
derive another empirical formula but rather exploit CR physics at
non-relativistic shocks.  Since the CR population is adiabatically compressed
at the shock in the limit of strong shocks, the total pressure jump is nearly
solely determined by the jump of the thermal pressure in the post-shock regime,
i.e.  we can safely neglect the contribution of CRs to the pressure jump.
This assumption is justified as long as the CR pressure is not dominated by
sub-relativistic CRs of low energy which is on the other hand a very short
lived population owing to Coulomb interactions in the ICM.  Thus, the thermal
post-shock pressure for $\M_1 \gtrsim 5$ is estimated as
\begin{equation}
  \label{eq:Pcorr}
  P_\rmn{th2} \simeq \frac{2\gamma_\rmn{th} \M_1^2 -
  (\gamma_\rmn{th}-1)} 
{\gamma_\rmn{th}+1}\, P_1,
\end{equation}
where $\M_1$ is obtained by equation~(\ref{eq:calibration}), and $P_1$ denotes the
stored total pre-shock pressure.

\section{Shock tubes}
\label{sec:shocktube}

\begin{figure*}
\begin{tabular}{cc}
{\large Polytropic thermal gas:} & {\large Composite of CRs and thermal gas:} \\
\resizebox{0.5\hsize}{!}{\includegraphics{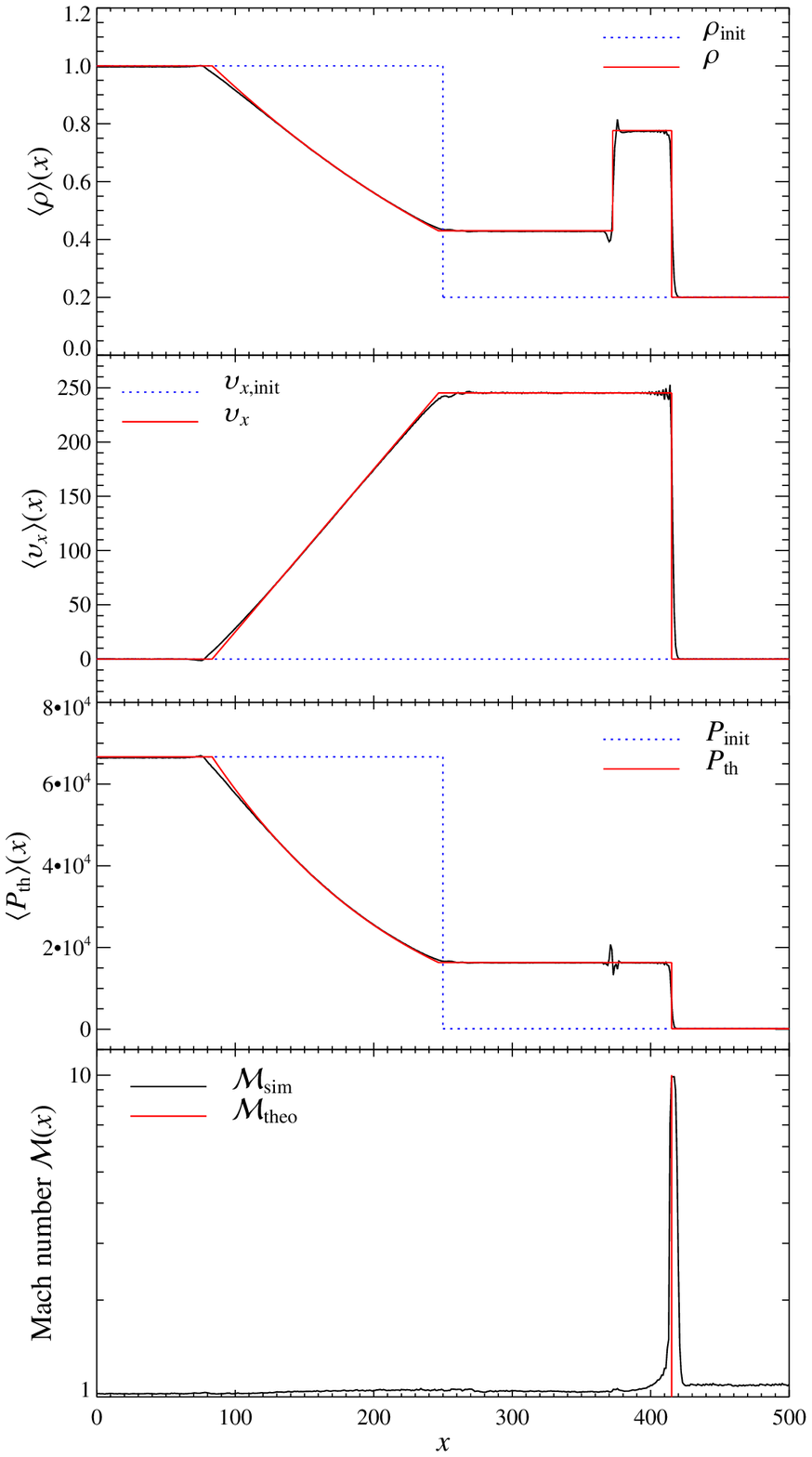}} &
\resizebox{0.5\hsize}{!}{\includegraphics{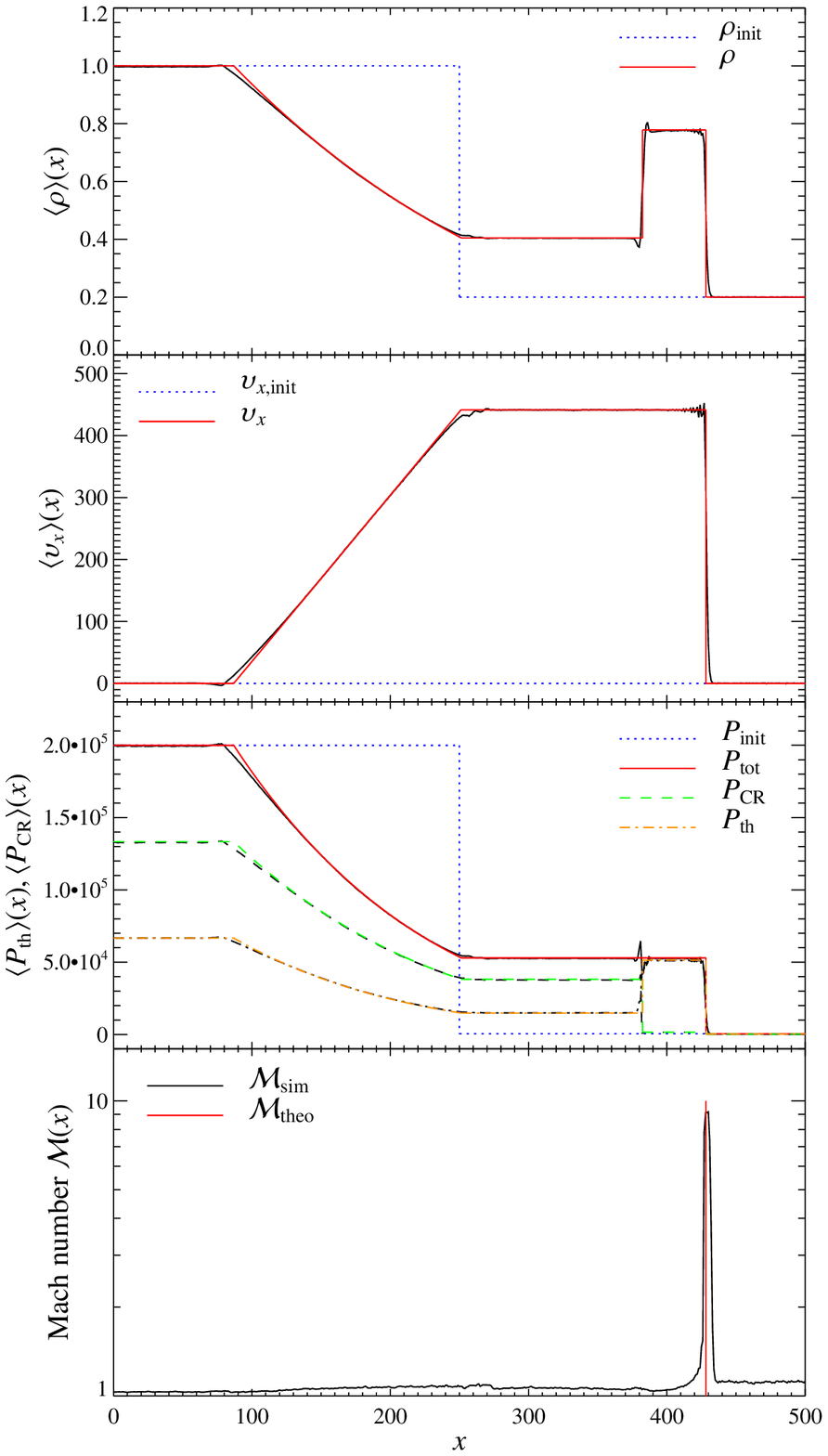}}
\end{tabular}
\caption{Shock-tube test carried out in a periodic three-dimensional box
  which is longer in $x$-direction than in the other two dimensions where a
  shock with the Mach number $\M=10$ develops.  The numerical result of the
  volume averaged hydrodynamical quantities $\bra \rho(x)\ket$, $\bra
  P(x)\ket$, $\bra \vel_x(x)\ket$, and $\bra \M(x)\ket$ within bins with a
  spacing equal to the interparticle separation of the denser medium is shown
  in black and compared with the analytic result in colour.  {\em Left-hand panels:}
  Shock-tubes are filled with pure thermal gas ($\gamma = 5/3$). {\em Right-hand
    panels:} Shock-tubes are filled with a composite of cosmic rays and thermal
  gas.  Initially, the relative CR pressure is $X_\CR = P_\CR / P_\th = 2$ in
  the left-half space ($x<250$), while we assume pressure equilibrium between
  CRs and thermal gas for $x>250$.}
\label{fig:shocktubes}
\end{figure*}

\begin{figure*}
\begin{tabular}{cc}
{\large Polytropic thermal gas:} & {\large Composite of CRs and thermal gas:} \\
\resizebox{0.5\hsize}{!}{\includegraphics{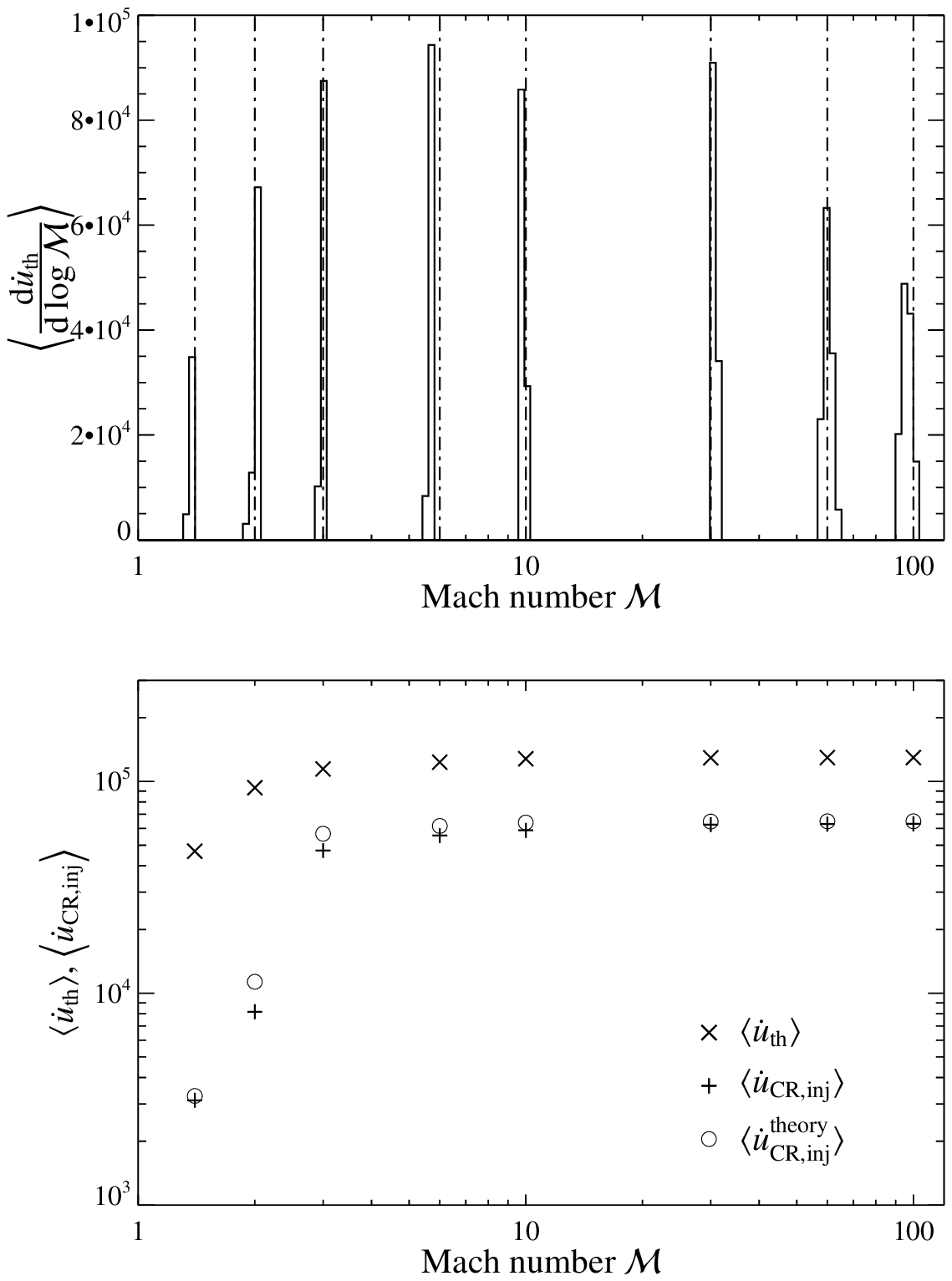}} &
\resizebox{0.5\hsize}{!}{\includegraphics{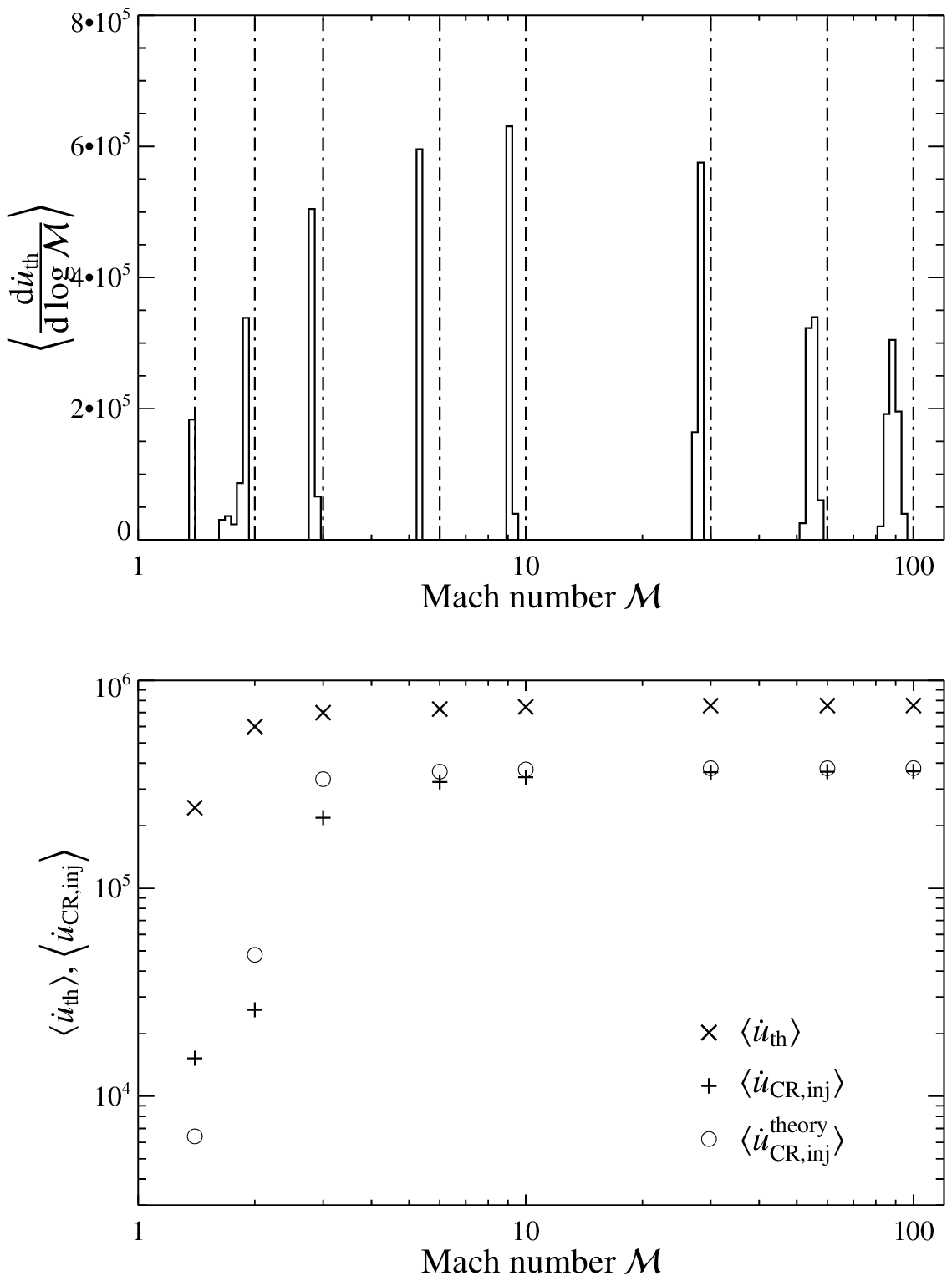}}
\end{tabular}
\caption{Mach number distributions weighted by the change of dissipated energy
  per time interval, $\left\bra \dd\dot{u}_\rmn{th} / (\dd \log\M)\right\ket$
  for our eight three-dimensional shock-tubes. {\em Left-hand panels:}
  Shock-tubes are filled with pure thermal gas ($\gamma = 5/3$). {\em
    Right-hand panels:} Shock-tubes are filled with a composite of cosmic rays
  and thermal gas.  Initially, the relative CR pressure is $X_\CR = P_\CR /
  P_\th = 2$ in the left-half space, while we assume pressure equilibrium
  between CRs and thermal gas.  {\em Bottom panels:} Shown are the change of
  dissipated energy per time interval, $\left\bra \dot{u}_\rmn{th} \right\ket$
  (shown with $\times$), the shock-injected CR energy $\bra
  \dot{u}_\rmn{CR,inj} \ket$ ($+$), and the theoretically expected injected CR
  energy $\bra \dot{u}_\rmn{CR,inj}^\rmn{theory} \ket$ ($\circ$) which is
  calculated following \citet{2005...Ensslin}.}
\label{fig:tubestatistics}
\end{figure*}

To assess the reliability of our formalism and the validity of our numerical
implementation, we perform a sequence of shock-tube simulations with Mach
numbers ranging from $\M=1.4$ up to $\M=100$. We use a three-dimensional
problem setup which is more demanding and more realistic than carrying out the
computation in one dimension.  Here and in the following, we drop the
subscript `1' of the pre-shock Mach number for convenience.  By comparing with
known analytic solutions, we are able to demonstrate the validity of our
implemented formalism.

There exists an analytic solution of the Riemann shock-tube problem in the
case of a fluid described by a polytropic equation of state, $\eps =
P/(\gamma-1)$ (cf.{\ }Appendix~\ref{sec:Riemann}). Unfortunately, a composite of
thermal gas and CRs does not obey this relation. Thus, we derive an analytic
solution to the Riemann shock-tube problem for the composite of CRs and thermal
gas in Appendix~\ref{sec:Riemann+CRs}.  This analytic solution assumes the CR
adiabatic index (equation~\ref{eq:gammaCR}) to be constant over the shock-tube
and neglects CR diffusion such that the problem remains analytically
treatable.

\subsection{Polytropic thermal gas}
\label{sec:shocktube_th}

We consider eight standard shock-tube tests \citep{1978JCoPh..27....1S} which
provide a validation of both the code's solution to hydrodynamic problems and
our Mach number formalism. We consider first an ideal gas with $\gamma=5/3$,
initially at rest. The left-half space ($x<250$) is filled with gas at unit
density, $\rho_2 = 1$, and $P_2 = (\gamma - 1)\, 10^5$, while $x>250$ is filled
with low density gas $\rho_1 = 0.2$ at low pressure. The exact value of the low
pressure gas has been chosen such that the resulting solutions yield the Mach
numbers $\M = \{1.4,2,3,6,10,30,60,100\}$ (cf.  Appendix~\ref{sec:Riemann}). We
set up the initial conditions in 3D using an irregular glass-like distribution
of a total of $3\times 10^4$ particles of equal mass in hydrostatic
equilibrium. They are contained in a periodic box which is longer in
$x$-direction than in the other two dimensions, $y$ and $z$.

In the left-hand panel of Fig.~\ref{fig:shocktubes}, we show the result for the
case of the Mach number $\M=10$ obtained with the \GADGET-2 code
\citep{2005MNRAS.364.1105S, 2001NewA....6...79S} at time $t = 0.5$.  Shown are
the volume averaged hydrodynamical quantities $\bra \rho(x)\ket$, $\bra
P(x)\ket$, $\bra \vel_x(x)\ket$, and $\bra \M(x)\ket$ within bins with a
spacing equal to the interparticle separation of the denser medium and
represented by solid black lines. One can clearly distinguish five regions of
gas with different hydrodynamical states.  These regions are separated by the
head and the tail of the leftwards propagating rarefaction wave, and the
rightwards propagating contact discontinuity and the shock wave.  The overall
agreement with the analytic solution is good, while the discontinuities are
resolved within $2-3$ SPH smoothing lengths. Despite the shock broadening, the
post-shock quantities are calculated very accurately.  Our formalism is clearly
able to detect the shock and precisely measure its strength, i.e.{\ }the Mach
number $\M$. The pressure quantity drawn is not the hydrodynamically acting
pressure of the SPH dynamics but $P = (\gamma-1) \rho u$, a product of two
fields that are calculated each using SPH interpolation.  Thus, the observed
characteristic pressure blip at the contact discontinuity has no real analogue
either in the averaged $x$-component of the velocity $\bra \vel_x(x) \ket$ or
in the averaged Mach number $\bra \M(x)\ket$. The $x$-component of the velocity
$\bra \vel_x(x) \ket$ shows tiny post-shock oscillations which might be damped
with higher values of the artificial viscosity in the expense of a broader
shock surface. The leftwards propagating rarefaction wave seems to exhibit a
slightly faster signal velocity compared to the sound velocity. This might be
attributed to the SPH averaging process which obtains additional information on
the SPH smoothing scale.

In the left-hand panel of Fig.~\ref{fig:tubestatistics}, we show the Mach number
distributions weighted by the change of dissipated energy per time interval,
$\left\bra \dd\dot{u}_\rmn{th} / (\dd \log\M)\right\ket$ for our eight
shock-tubes. The sharp peaks of these distributions around their expected
values $\log \M$ are apparent. This demonstrates the reliability of our
formalism to precisely measure shock strengths instantaneously during SPH
simulations.  The bottom panel shows their integral, i.e.{\ }the change of
dissipated energy per time interval, $\left\bra \dot{u}_\rmn{th} \right\ket$.
The rising dissipated energy with growing Mach number reflects the larger
amount of available kinetic energy for dissipation.

We additionally calculate the shock-injected CR energy using our formalism of
diffusive shock acceleration described in \citet{2005...Ensslin}. However, the
injected CR energy $\left\bra \dot{u}_\rmn{CR,inj} \right\ket$ was only
monitored and not dynamically tracked. For comparison, we also show the
theoretically expected injected CR energy $\left\bra
  \dot{u}_\rmn{CR,inj}^\rmn{theory} \right\ket = \zeta_\rmn{inj} \left\bra
  \dot{u}_\rmn{th} \right\ket$, where $\zeta_\rmn{inj}$ is the energy
efficiency due to diffusive shock acceleration \citep[cf.][for
details]{2005...Ensslin}.  The good comparison of the simulated and
theoretically expected shock-injected CR energy demonstrates that our
formalism is reliably able to describe the on-the-fly acceleration of CRs
during the simulation.

\subsection{Composite of cosmic rays and thermal gas}

Again, we consider eight shock-tube simulations containing a composite
of cosmic rays and thermal gas, providing a useful validation of our
CR implementation in solving basic hydrodynamic problems as well as
our Mach number formalism in the presence of CRs. In these
simulations, we neither inject shock-accelerated CRs nor consider CR
diffusion: these processes would lead to CR modified shock structures
and shall be the subject of a companion paper.

To characterise this composite fluid, we define the relative CR
pressure $X_\CR = P_\CR / P_\th$. Our composite gas is initially at
rest, while the left-half space ($x<250$) is filled with gas at unit
density, $\rho_2 = 1$, $X_{\CR2} = 2$, and $P_\rmn{th2} = (\gamma -
1)\, 10^5$, while $x>250$ is filled with low density gas $\rho_1 =
0.2$, $X_{\CR1} = 1$, at low pressure.  The exact value of the low
pressure gas has again been chosen such that the resulting solutions
yield the Mach numbers $\M = \{1.4,2,3,6,10,30,60,100\}$ (cf.
Appendix~\ref{sec:Riemann+CRs}).  Otherwise, we use the same initial
setup as in Section~\ref{sec:shocktube_th}. This CR load represents a
rather extreme case and can be taken as the limiting case for our Mach
number formalism in the presence of CRs. Cosmologically, it may find
application in galaxy mergers where the outer regions might be
composed of an adiabatically expanded composite gas containing a high
CR component.

In the right-hand panel of Fig.~\ref{fig:shocktubes}, we show the result for the
case of the Mach number $\M=10$ obtained with \GADGET-2 at time $t = 0.3$.  The
agreement with the analytic solution is good, while the discontinuities are
resolved within $2-3$ SPH smoothing lengths. Despite the shock broadening, the
post-shock quantities are calculated very accurately.  In the case of composite
gas, our formalism is clearly able to detect the shock and measure its
strength with a Mach number accuracy better than 10\%. Although the total
pressure remains constant across the contact discontinuity, the partial
pressure of CRs and thermal gas interestingly are changing. This behaviour
reflects the adiabatic compression of the CR pressure component across the
shock wave. A posteriori, this justifies our procedure of inferring the thermal
pressure jump at the shock for a composite of CRs and thermal gas in
equation~(\ref{eq:Pcorr}).

In the right-hand panel of Fig.~\ref{fig:tubestatistics}, we show the Mach number
distributions weighted by the change of dissipated energy per time interval,
$\left\bra \dd\dot{u}_\rmn{th} / (\dd \log\M)\right\ket$ for our eight
shock-tubes. While our formalism is able to measure the shock strength with a
Mach number accuracy better than 10\%, the distributions are sharply peaked.
This demonstrates the reliability of our formalism to measure shock strengths
for the composite gas instantaneously during SPH simulations. 

The bottom panel shows the change of dissipated energy per time
interval, $\left\bra \dot{u}_\rmn{th} \right\ket$ together with the
shock-injected CR energy $\left\bra \dot{u}_\rmn{CR,inj} \right\ket$.
Concerning the amount of injected CR energy, we neglected cooling
processes such as Coulomb interactions with thermal particles: this
would effectively result in a density dependent recalibration of the
maximum CR energy efficiency $\zeta_\rmn{max}$ of the otherwise
arbitrary absolute value of our fiducial density. In the case of high
Mach numbers, there is a good agreement between the simulated and
theoretically expected shock-injected CR energy while there are
discrepancies at low Mach numbers: our formalism estimates volume
averaged Mach numbers with an accuracy better than 10\%; this
uncertainty translates to estimates of the density jump $x_\s$ and the
thermal pressure jump $y_\s$ with a scatter among different SPH
particles. In the regime of weak shocks, the CR energy efficiency due
to diffusive shock acceleration $\zeta_\rmn{inj}$ is extremely
sensitive to these two quantities, leading to larger uncertainties for
the shock-injected CR energy in the case of a high CR load. However,
the overall trend for the shock-injected CR energy can still be
matched in such an extreme physical environment.

\section{Non-radiative cosmological simulations}
\label{sec:simulations} 

\subsection{Simulation setup}

As a first application of our formalism, we are here interested in studying the
spatial distribution of cosmological structure formation shocks in combination
with Mach number statistics. We focus on the ``concordance'' cosmological cold
dark matter model with a cosmological constant ($\Lambda$CDM). The cosmological
parameters of our model are: $\Omega_\rmn{m} = \Omega_\rmn{dm} + \Omega_\rmn{b}
= 0.3$, $\Omega_\rmn{b} = 0.04$, $\Omega_\Lambda = 0.7$, $h = 0.7$, $n = 1$,
and $\sigma_8 = 0.9$. Here, $\Omega_\rmn{m}$ denotes the total matter density
in units of the critical density for geometrical closure, $\rho_\rmn{crit} = 3
H_0^2 / (8 \upi G)$. $\Omega_\rmn{b}$ and $\Omega_\Lambda$ denote the densities
of baryons and the cosmological constant at the present day. The Hubble
constant at the present day is parametrized as $H_0 = 100\,h \mbox{ km s}^{-1}
\mbox{Mpc}^{-1}$, while $n$ denotes the spectral index of the primordial
power-spectrum, and $\sigma_8$ is the {\em rms} linear mass fluctuation within
a sphere of radius $8\,h^{-1}$Mpc extrapolated to $z=0$. This model yields a
reasonable fit to current cosmological constraints and provides a good
framework for investigating cosmological shocks.

Our simulations were carried out with an updated and extended version of the
distributed-memory parallel TreeSPH code \GADGET-2 \citep{2005MNRAS.364.1105S,
  2001NewA....6...79S} including now self-consistent cosmic ray physics
\citep{2005...Ensslin,2005...Jubelgas}. Our reference simulation employed
$2\times 256^3$ particles which were simulated within a periodic box of
comoving size $100\,h^{-1}$Mpc, so the dark matter particles had masses of
$4.3 \times 10^9\,h^{-1}\,M_\odot$ and the SPH particles $6.6\times
10^8\,h^{-1}\,M_\odot$.  The SPH densities were computed from 32 neighbours
which leads to our minimum gas resolution of approximately $2\times
10^{10}\,h^{-1}\,M_\odot$.  The gravitational force softening was of a spline
form \citep[e.g.,][]{1989ApJS...70..419H} with a Plummer-equivalent softening
length of $13\,h^{-1}$kpc comoving.  In order to test our numerical
resolution, we additionally simulated the same cosmological model with
$2\times 128^3$ particles, with a softening length twice that of the reference
simulation.

Initial conditions were laid down by perturbing a homogeneous particle
distribution with a realization of a Gaussian random field with the
$\Lambda$CDM linear power spectrum. The displacement field in Fourier
space was constructed using the Zel'dovich approximation, with the
amplitude of each random phase mode drawn from a Rayleigh
distribution.  For the initial redshift we chose $1+z_\rmn{init}=50$
which translates to an initial temperature of the gas of
$T_\rmn{init}=57$~K.  This reflects the fact that the baryons are
thermally coupled to the CMB photons via Compton interactions with the
residual free electrons after the universe became transparent until it
eventually decoupled at $1+z_\rmn{dec} \simeq 100 (\Omega_\rmn{b} h^2
/ 0.0125)^{2/5}$.  In all our simulations, we stored the full particle
data at 100 output times, equally spaced in $\log (1+z)$ between
$z=40$ and $z=0$.

In order to investigate the effects of reionisation on the Mach number
statistics, we additionally perform two similar simulations which
contain a simple reionisation model where we impose a minimum gas
temperature of $T=10^4$~K at a redshift of $z=10$ to all SPH
particles. We decided to adopt this simplified model to study its
effect on the Mach number statistics rather than a more complicated
reionisation history. A more realistic scenario might be to add energy
only to gas within haloes above a certain density in combination with
energy input from QSO activity, and to describe the merging of the
reionisation fronts and their evolution into the lower density regions
\citep[e.g.,][]{2003MNRAS.343.1101C}.

The simulation reported here follow only non-radiative gas physics. We
neglected several physical processes, such as radiative cooling,
galaxy/star formation, and feedback from galaxies and stars including
cosmic ray pressure. Our primary focus are shocks that are mostly
outside the cluster core regions. Thus, the conclusions drawn in this
work should not be significantly weakened by the exclusion of these
additional radiative processes.

In contrast to the idealised shock tube experiment where all particles that are
shocked experience the same shock strength, in cosmological simulations there
might be a a distribution of Mach numbers for a given region of space because
of curvature effects, multiple shocks, etc. Thus the Mach number estimation in
shock tubes involves averaging over many particles all of which are
experiencing a shock of a given Mach number, whereas in the cosmological
simulations the averaging has to be done over Mach number also. This might
introduce a scatter to the Mach number estimation of a single particle in
cosmological simulations which is difficult to quantify. We are confident that
this effect has only a minor impact on our results because they agree well with
results of similar studies that used Eulerian structure formation simulations
\citep{2003ApJ...593..599R}.

\begin{figure*}
\resizebox{\hsize}{!}{\includegraphics{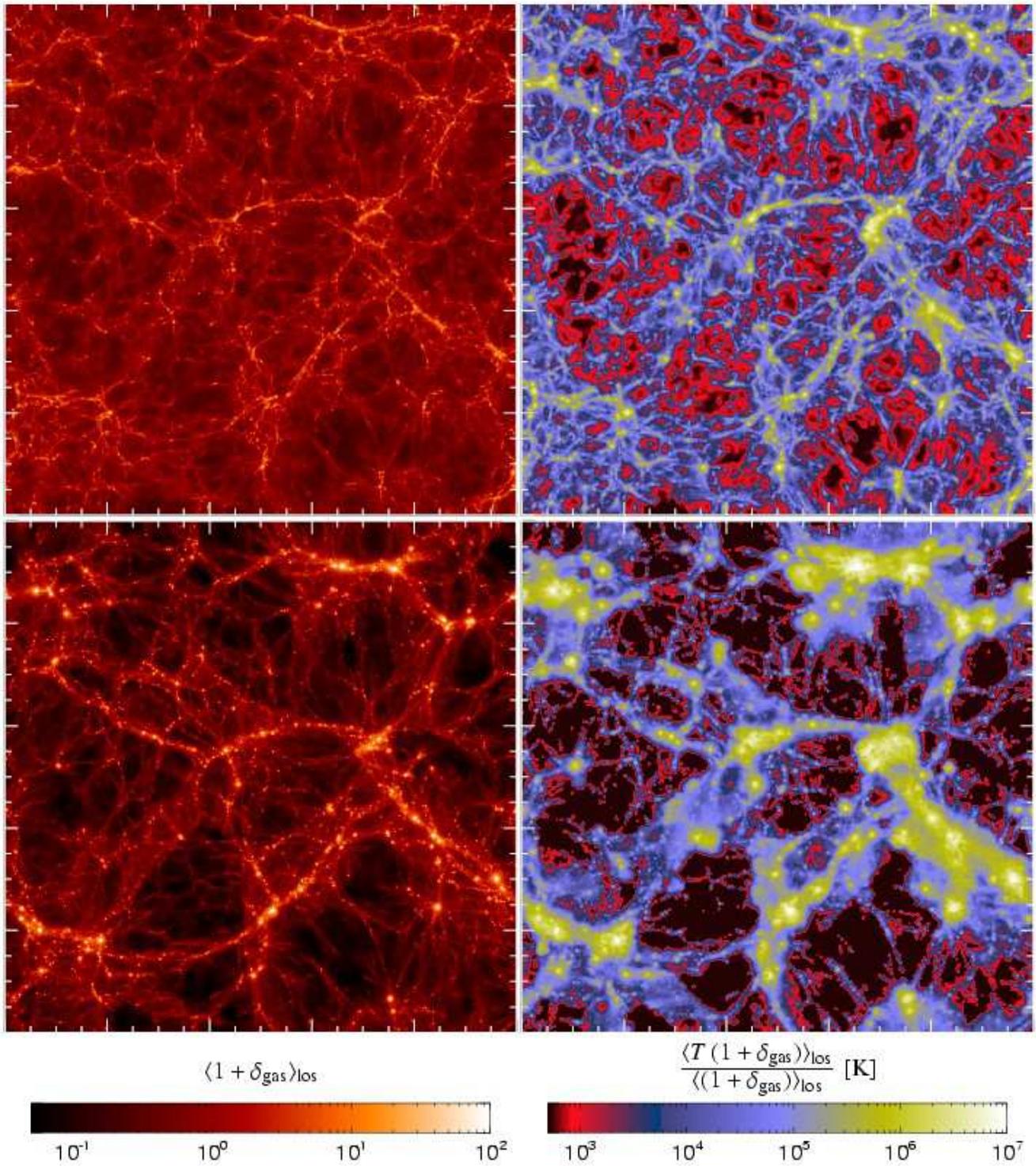}}
\caption{Visualisation of a non-radiative cosmological simulation at redshift
  $z=2$ ({\em top panels}) and $z=0$ ({\em bottom panels}). Shown are the
    overdensity of the gas ({\em left-hand side}) and the density weighed gas
    temperature ({\em right-hand side}). These pictures have a comoving side
    length of $100~h^{-1}\,$Mpc while the projection length along the
    line-of-sight amounts to $10~h^{-1}\,$Mpc. }
\label{fig:paneldens}
\end{figure*}

\begin{figure*}
\resizebox{\hsize}{!}{\includegraphics{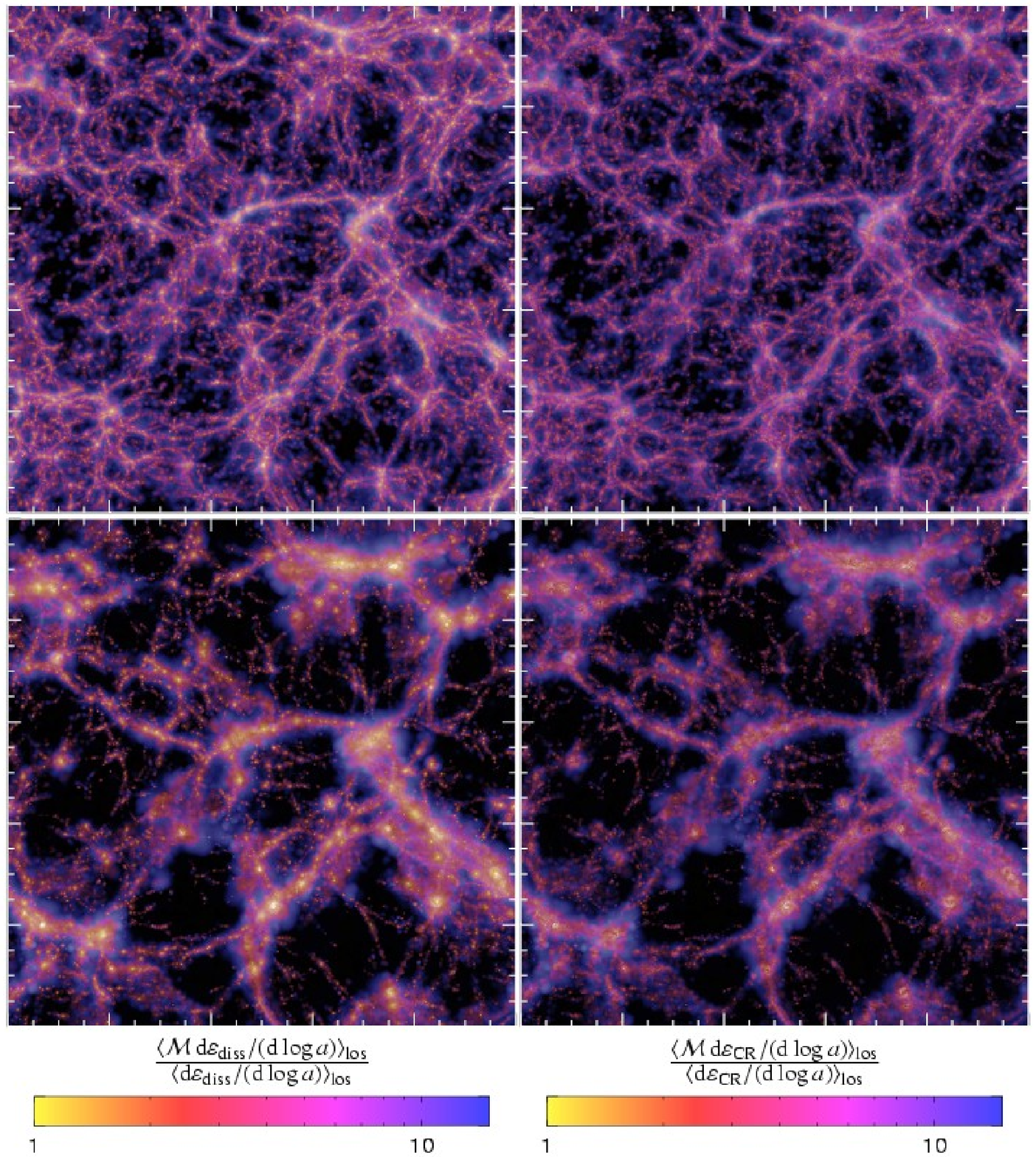}}
\caption{Mach number visualisation of a non-radiative cosmological
  simulation at redshift $z=2$ ({\em top panels}) and $z=0$ ({\em
  bottom panels}). The colour hue of the maps on the left-hand side
  encodes the spatial Mach number distribution weighted by the rate of
  energy dissipation at the shocks, normalised to the simulation
  volume. The maps on the right-hand side show instead the Mach number
  distribution weighted by the rate of CR energy injection above
  $q=0.8$, the threshold of hadronic interactions.  The brightness of
  each pixel is determined by the respective weights, i.e.~by the
  energy production density.  These pictures have a comoving length of
  $100~h^{-1}\,$Mpc on a side. Most of the energy is dissipated in
  weak shocks which are situated in the internal regions of groups or
  clusters, while collapsed cosmological structures are surrounded by
  strong external shocks (shown in blue). }
\label{fig:panelmach} 
\end{figure*}

\subsection{Visualisation of the Mach number}

In the SPH formalism, continuous fields $A(\vecbf{x})$ such as the
hydrodynamical quantities are represented by the values $A_i$ at discrete
particle positions $\vecbf{r}_i = (x_i,y_i,z_i)$ with a local spatial
resolution given by the SPH smoothing length $h_i$. To visualise a scalar
quantity in two dimensions we employ the mass conserving {\em scatter} approach
for the projection, where the particle's smoothing kernel is distributed onto
cells of a Cartesian grid which is characterised by its physical mesh size $g$.
The line-of-sight integration of any quantity $A(\vecbf{x})$ at the pixel at
position $\vecbf{r} = (x,y,z)$ is determined as the average of integration of
all lines of sight passing through the pixel,
\begin{equation}
  \label{eq:SPHprojection}
  \left\bra a(\vecbf{x}_\bot)\right\ket_\rmn{los} = 
  g^{-2}\! \sum_i h_i^{-3}
  \left[\int_{x-g/2}^{x+g/2}\!\!\dd x_i\!
        \int_{y-g/2}^{y+g/2}\!\!\dd y_i\!
        \int_{-h_i}^{h_i}\!\!\dd z_i \,\mathcal{K}\left(\frac{r}{h_i}\right) 
        A_i\right],
\end{equation}
with $r = \sqrt{(x_i-x)^2 + (y_i-y)^2 + z_i^2}$, and where the summation is
extended over all particles in the desired slice of projection. The function
$\mathcal{K}$ is the dimensionless spherically symmetric cubic spline kernel
suggested by \citet{1985A&A...149..135M}.

The left-hand side of Fig.~\ref{fig:paneldens} shows the time evolution of the
density contrast $\delta$ averaged over the line-of-sight with a comoving
projection length $L_\rmn{proj} = 10~h^{-1}\,$Mpc:
\begin{equation}
  \label{eq:1+delta}
  \left\bra 1 + \delta_\rmn{gas}(\vecbf{x}_\bot)\right\ket_\rmn{los} = 
  \frac{\left\bra \Sigma(\vecbf{x}_\bot)\right\ket_\rmn{los}}
       {L_\rmn{proj}\,\rho_\rmn{crit}\,\Omega_\rmn{b}},
\end{equation}
where $\Sigma$ denotes the surface mass density. The fine-spun cosmic
web at high redshift evolves into a much more pronounced, knotty and
filamentary structure at late times, as a result of the hierarchical
structure formation process driven by gravity.

The right-hand side of Fig.~\ref{fig:paneldens} shows the time evolution of the
density weighted temperature averaged over the line-of-sight. Again, the growth
of galaxy clusters visible as large bright regions with temperatures around
$10^7$K is clearly visible. Through dissipation, the shock waves convert part
of the gravitational energy associated with cosmological structure formation
into internal energy of the gas, apart from the additional contribution due to
adiabatic compression caused by the material that falls in at later times and
itself is compressed at these shock waves.  The large black regions show voids
which cool down during cosmic evolution due to two effects: while the universe
expands, non-relativistic gas is adiabatically expanded and cools according to
$T\propto V^{1-\gamma}\propto a^{-2}$ for $\gamma=5/3$ when shock heating is
still absent. Secondly, matter is flowing towards filaments during structure
formation, hence the voids get depleted, providing an additional adiabatic
expansion of the remaining material.

Fig.~\ref{fig:panelmach} shows a visualisation of the responsible
structure formation shocks and their corresponding strengths.  The
colour scaling represents the spatial Mach number distribution weighted
by the rate of energy dissipation at the shocks, and normalised to the
simulation volume (left-hand side). The Mach number distribution
weighted by the rate of CR energy injection is shown in the right-hand
side, again normalised to the simulation volume. The brightness of
these pixels scales with the respective weights, i.e.~by the rates of
energy dissipation or injection, respectively. The spatial Mach number
distribution impressively reflects the nonlinear structures and voids
of the density and temperature maps of Fig.~\ref{fig:paneldens}.  It
is apparent that most of the energy is dissipated in weak shocks which
are situated in the internal regions of groups or clusters while
collapsed cosmological structures are surrounded by external strong
shocks (shown in blue).  These external shocks are often referred to
as `first shocks', because here the compressed gas has been processed for
the first time in its cosmic history through shock waves.

Following \citet{2003ApJ...593..599R}, we classify structure formation shocks
into two broad populations which are labelled as {\em internal} and {\em
  external} shocks, depending on whether or not the associated pre-shock gas
was previously shocked. Rather than using a thermodynamical criterion such as
the temperature, we prefer a criterion such as the overdensity $\delta$ in
order not to confuse the shock definition once we will follow radiatively
cooling gas in galaxies (in practice, we use the criterion of a critical
pre-shock overdensity $\delta>10$ for the classification of an internal shock).
{\em External} shocks surround filaments, sheets, and haloes while {\em
  internal} shocks are located within the regions bound by external shocks and
are created by flow motions accompanying hierarchical structure formation.  For
more detailed studies, internal shocks can be further divided into three types
of shock waves: (1) accretion shocks caused by infalling gas from sheets to filaments
or haloes and from filaments to haloes, (2) merger shocks resulting from merging
haloes, and (3) supersonic chaotic flow shocks inside nonlinear structures which
are produced in the course of hierarchical clustering.

In contrast to the present time, the comoving surface area of external
shock waves surpasses that of internal shocks at high redshift, due to
the small fraction of mass bound in large haloes and the simultaneous
existence of an all pervading fine-spun cosmic web with large surface
area.  Also, there the thermal gas has a low sound velocity $c =
\sqrt{\gamma P/\rho}=\sqrt{\gamma (\gamma -1) u}$ owing to the low
temperature, so once the diffuse gas breaks on mildly nonlinear
structures, strong shock waves develop that are characterised by high
Mach numbers $\M=\vel_\rmn{s}/ c$.  Nevertheless, the energy
dissipation rate in internal shocks is always higher compared to
external shocks because the mean shock speed and pre-shock gas
densities are significantly larger for internal shocks.

\begin{figure}
\resizebox{\hsize}{!}{\includegraphics{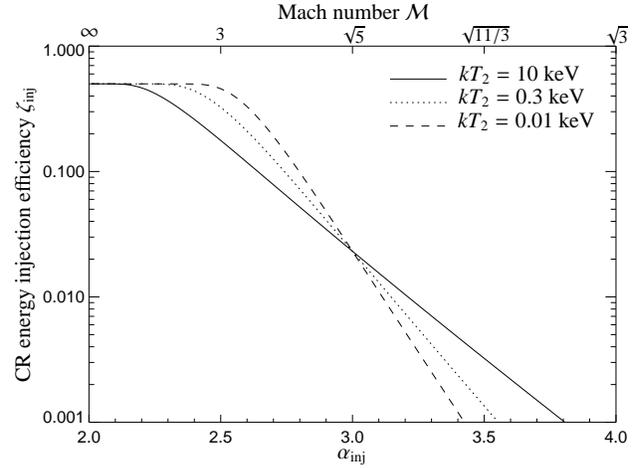}}
\caption{CR energy injection efficiency for the diffusive shock
  acceleration process. Shown is the CR energy injection efficiency
  $\zeta_\rmn{inj} = \eps_\rmn{CR}/\eps_\rmn{diss}$ for the three
  post-shock temperatures $kT_2/\mbox{keV} = 0.01, 0.3$, and 10. We
  inject only CRs above the kinematic threshold $q_\rmn{threshold} =
  0.83$ of the hadronic CRp-p interaction which are able to produce
  pions that successively decay into secondary electrons, neutrinos,
  and $\gamma$-rays. We choose equipartition between injected CR
  energy and dissipated thermal non-relativistic energy as saturation
  value of the CR energy injection efficiency, i.e. $\zeta_\rmn{max} =
  0.5$ \citep{2003ApJ...593..599R}.}
\label{fig:zeta}
\end{figure}

We use the same colour and brightness scale for the Mach number
distribution weighted by the injected CR energy rate normalised to the
simulation volume (right-hand side of Fig.~\ref{fig:panelmach}). We
emphasize two important points which have fundamental implications for
the CR population in galaxy clusters. (1) There is an absence of weak
shocks (shown in yellow) when the Mach number distribution is weighted
by the injected CR energy. This reflects the Mach number dependent
energy injection efficiency: the CR injection is saturated for strong
shocks which leads to similar spatial distribution of both weightings,
by dissipated energy as well as by injected CR energy. In contrast,
most of the dissipated energy is thermalized in weak shocks and only
small parts are used for the acceleration of relativistic particles
(compare Fig.~\ref{fig:zeta}).  (2) The mechanism of energy
dissipation at shocks is very density dependent, implying a tight
correlation of weak internal shocks and the amount of dissipated
energy.  This can be seen by the strongly peaked brightness
distribution of the dissipated energy rate towards the cluster centres.
For the CR-weighted case, this correlation is counteracted by the CR
energy injection efficiency leading to a smoother brightness
distribution of the CR energy injection. This has the important
implication that the ratio of CR injected energy to dissipated thermal
energy is increasing as the density declines.  Relative to
the thermal non-relativistic energy density, the CR energy density is
dynamically more important at the outer cluster regions and
dynamically less important at the cluster centres.

\subsection{Mach number statistics}

\subsubsection{Influence of reionisation}

\begin{figure*}
\resizebox{\hsize}{!}{\includegraphics{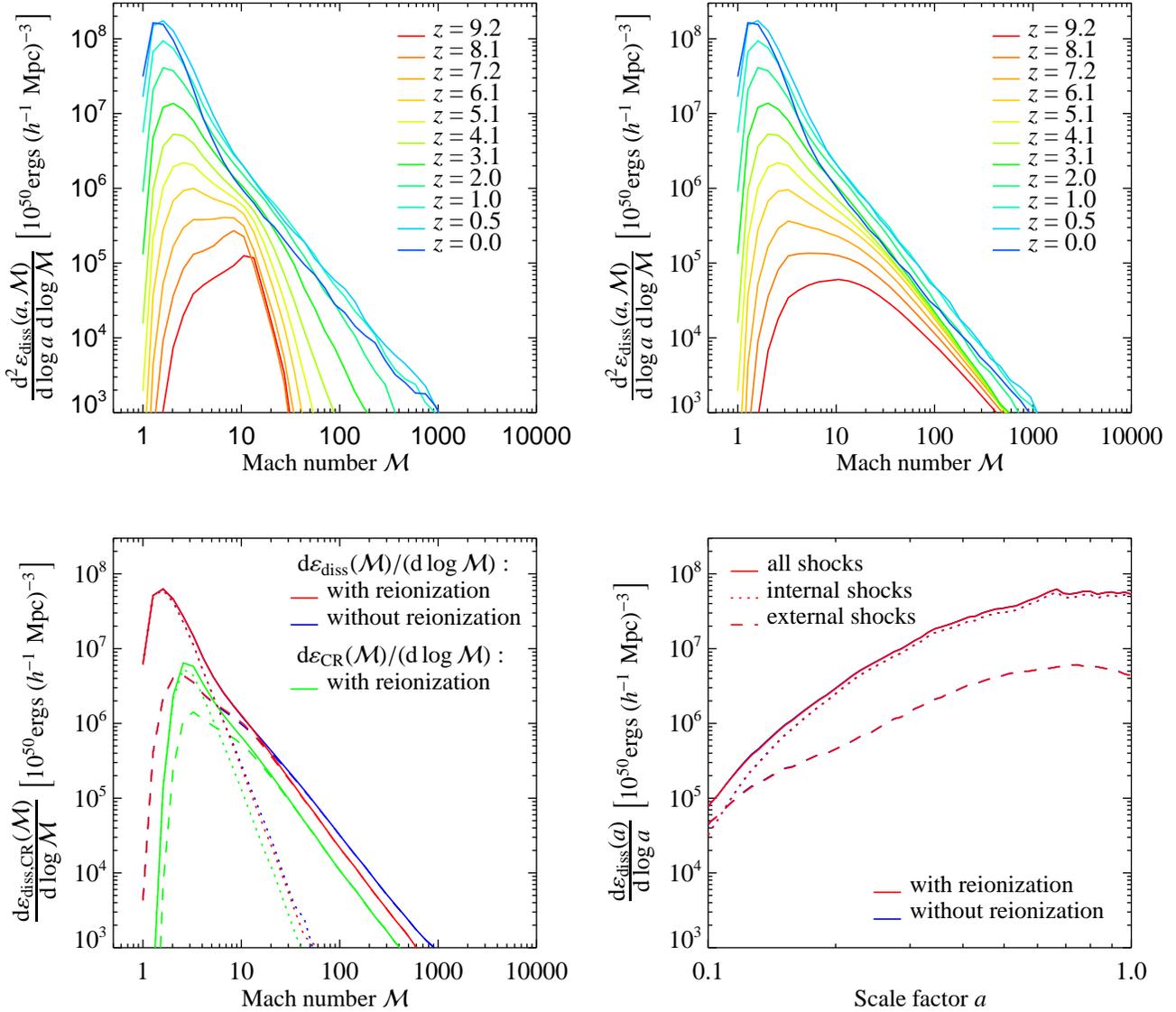}}
\caption{Influence of reionisation on the Mach number statistics of
  non-radiative cosmological simulations. The {\em top left-hand panel} shows the
  differential Mach number distribution $\dd^2 \eps_\rmn{diss}(a,\M)/( \dd \log
  a\, \dd \log\M)$ for our reference simulation with reionisation while the
  {\em top right-hand panel} shows this distribution for the simulation without
  reionisation.  The {\em lower left-hand panel} shows both distributions integrated
  over the scale factor, $\dd \eps_\rmn{diss}(\M)/( \dd \log \M)$ in addition
  to the Mach number distribution weighted by the injected CR energy rate
  normalised to the simulation volume, $\dd \eps_\rmn{CR}(\M)/( \dd \log \M)$
  (green). Internal shocks are shown with dotted lines and external shocks with
  dashed lines. The {\em lower right-hand panel} shows the evolution of the
  dissipated energy due to shock waves with scale factor, $\dd
  \eps_\rmn{diss}(a)/( \dd \log a)$. The models with and without reionisation
  lie on top of each other.}
\label{fig:reionization}
\end{figure*}

To quantify previous considerations, we compute the differential Mach number
distribution weighted by the dissipated energy normalised to the simulation
volume $\dd^2 \eps_\rmn{diss} (a,\M)/( \dd \log a\, \dd \log\M)$ at different
redshifts. The top left-hand panel of Fig.~\ref{fig:reionization} shows this Mach
number distribution for our reference simulation with reionisation (showing a
resolution of $2\times 256^3$), while the top right-hand panel shows this
distribution for the simulation without reionisation.  The lower left-hand panel
shows both distributions integrated over the scale factor, $\dd
\eps_\rmn{diss}(\M)/( \dd \log \M)$, in addition to the Mach number
distribution weighted by the injected CR energy normalised to the simulation
volume, $\dd \eps_\rmn{CR}(\M)/( \dd \log \M)$ (see
Section~\ref{sec:CRacceleration}).  Internal shocks are shown with dotted lines
and external shocks with dashed lines. The lower right-hand panel shows the
evolution of the dissipated energy due to shock waves with scale factor, $\dd
\eps_\rmn{diss}(a)/( \dd \log a)$, for the models with and without
reionisation.

Several important points are apparent. (1) The median of the Mach number
distribution weighted by the dissipated energy decreases as cosmic time
evolves, i.e.{\ }the average shock becomes weaker at later times.  (2) There is
an increasing amount of energy dissipated at shock waves as the universe
evolves because the mean shock speed is significantly growing when the
characteristic mass becomes larger with time.  This trend starts to level off
at redshift $z\simeq 1$ although the median Mach number in shocks continuous to
decrease. (3) Reionisation influences the Mach number distribution
predominantly at early times (however after reionisation took place) and
suppresses strong external shock waves efficiently. The reason is that
reionisation of the thermal gas increases its sound speed $c = \sqrt{\gamma n k
  T/\rho}$ dramatically, so that weaker shocks are produced for the same shock
velocities.  (4) The time integrated Mach number distribution weighted by the
dissipated energy peaks at Mach numbers $1\lesssim\M\lesssim 3$. The main
contribution in terms of energy dissipation originates from internal shocks
because of enhanced pre-shock densities and mean shock speeds. (5) External
shocks dominate the Mach number distribution at early times while internal
shocks take over at $z \simeq 9$ (depending somewhat on the resolution of the
simulation).  Their amount of dissipated energy surpasses that in external
shocks by over an order of magnitude at the present time. Internal shocks play
a more important role than external shocks in dissipating energy associated
with structure formation.

\begin{figure*}
  \resizebox{\hsize}{!}{\includegraphics{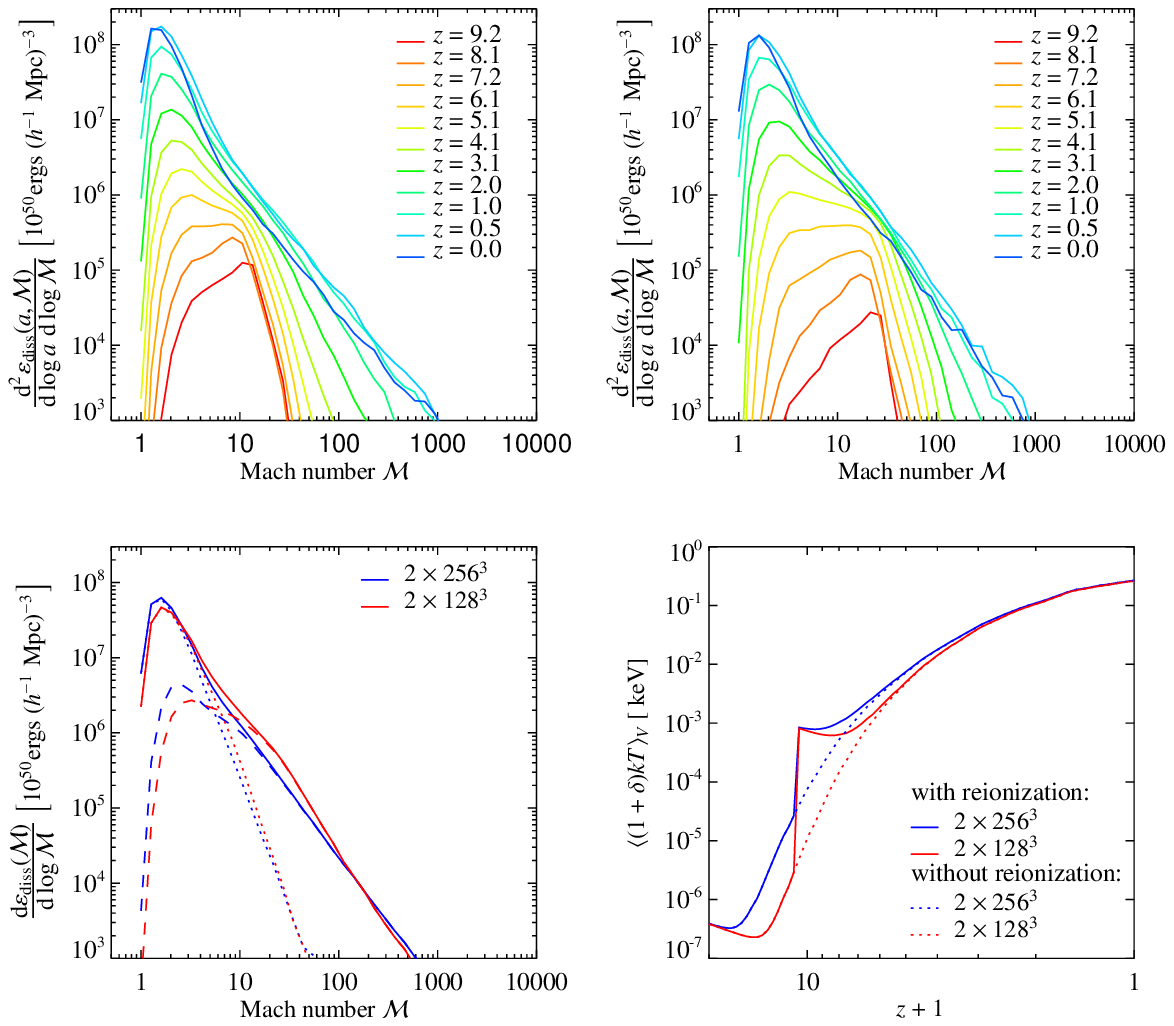}}
\caption{Resolution study: Mach number statistics for non-radiative
  cosmological simulations with a reionisation epoch at $z=10$. The {\em top
  left-hand panel} shows the differential Mach number distribution $\dd^2
  \eps_\rmn{diss}(a,\M)/( \dd \log a\, \dd \log\M)$ for our reference
  simulation with a resolution of $2\times 256^3$ while the {\em top right-hand
  panel} shows this distribution for the simulation with a resolution of
  $2\times 128^3$.  The {\em lower left-hand panel} shows both distributions
  integrated over the scale factor, $\dd \eps_\rmn{diss}(\M)/( \dd \log \M)$.
  Internal shocks are shown with dotted lines and external shocks with dashed
  lines. The {\em lower right-hand panel} shows the evolution of the
  density-weighted temperature with redshift. Shown are different resolutions
  in our models with and without reionisation.}
\label{fig:resolution}
\end{figure*}

The total shock-dissipated energy in our simulation box amounts to
$E_\rmn{diss} = 2.27\times 10^{64} \mbox{ erg}$. This translates to a mean
energy deposition per particle of $E_\rmn{diss}\, \mu / (\rho_\rmn{crit}
\Omega_\rmn{b} V) = 0.63 \mbox{ keV}$, where $\mu = 4 m_\rmn{p}/(3 + 5
X_\rmn{H})$ is the mean particle weight assuming full ionisation and
$X_\rmn{H}=0.76$ is the primordial hydrogen mass fraction. Our results agree
well with those of \citet{2003ApJ...593..599R} in the case of internal shocks
while our external shocks tend to be weaker. This can be attributed to our
differing definition of internal/external shocks as we prefer a density
criterion and use the critical pre-shock overdensity $\delta>10$ for the
classification of an internal shock.

\subsubsection{Cosmic ray acceleration}
\label{sec:CRacceleration}

In our non-radiative cosmological simulations we additionally calculate the
expected shock-injected CR energy using our formalism of diffusive shock
acceleration described in \citet{2005...Ensslin}.  This formalism follows a
model based on plasma physics for the leakage of thermal ions into the CR
population.  However, in the present analysis, the injected CR energy $\dd
u_\rmn{CR,inj} /(\dd \log a)$ was only monitored and not dynamically tracked.
In our model, the CR population is described by single power-law distribution
which is uniquely determined by the dimensionless momentum cutoff $q$, the
normalisation $C$, and the spectral index $\alpha$. Considering shock injected
CRs only, the spectral index is determined by $\alpha =
(x_\rmn{s}+2)/(x_\rmn{s}-1)$, where $x_\rmn{s}$ denotes the density jump at
the shock.

Our simplified model for the diffusive shock acceleration fails in the
limit of weak shocks and over-predicts the injection
efficiency. Especially in this regime, Coulomb losses have to be taken
into account which remove the low-energetic part of the injected CR
spectrum efficiently on a short timescale giving rise to an effective
CR energy efficiency. Thus, the instantaneous injected CR energy $\dd
u_\rmn{CR,inj} /(\dd \log a)$ depends on the simulation timestep and
the resolution. To provide a resolution independent statement about
the injected CR energy, we decided to rethermalise the injected CR
energy below the cutoff $q_\rmn{threshold} = 0.83$.  This cutoff has
the desired property, that it coincides with the kinematic threshold
of the hadronic CR p-p interaction to produce pions which decay into
secondary electrons (and neutrinos) and $\gamma$-rays:
\begin{eqnarray}
  \pi^\pm &\rightarrow& \mu^\pm + \nu_{\mu}/\bar{\nu}_{\mu} \rightarrow
  e^\pm + \nu_{e}/\bar{\nu}_{e} + \nu_{\mu} + \bar{\nu}_{\mu}\nonumber\\
  \pi^0 &\rightarrow& 2 \gamma \,.\nonumber
\end{eqnarray}
Only CR protons above this kinematic threshold are therefore visible through
their decay products in both the $\gamma$-ray and radio bands via radiative
processes, making them directly observationally detectable.

The lower left-hand panel of Fig.~\ref{fig:reionization} shows the Mach number
distribution weighted by the injected CR energy rate and normalised to the
simulation volume, $\dd \eps_\rmn{CR}(\M)/( \dd \log \M)$ (solid green).  The
effect of the CR injection efficiency $\zeta_\rmn{inj} = \eps_\rmn{CR} /
\eps_\rmn{diss}$ can easily be seen: while the CR injection is saturated for
strong shocks to $\zeta_\rmn{max} = 0.5$, in weak shocks most of the dissipated
energy is thermalized and only small parts are used for the acceleration of
relativistic particles. Effectively, this shifts the maximum and the mean value
of the Mach number distribution weighted by the shock-dissipated energy towards
higher values in the case of the distribution weighted by the injected CR
energy. This effect is even stronger when considering only CRs with a lower
cutoff $q=10,30$ which are responsible for radio haloes observed at frequencies
above $100$~MHz, assuming typical magnetic field strengths of $B=10,1\,\mu$G,
respectively.  This follows from the mono-energetic approximation of the
hadronic electron production and synchrotron formula,
\begin{equation}
  \label{eq:synchrotron}
  \nu_\rmn{s} = \frac{3 e B}{2\pi m_\rmn{e} c}\gamma_\rmn{e}^2,
  \quad\mbox{where}\quad
  \gamma_\rmn{e} \simeq \frac{q}{16}\frac{m_\rmn{p}}{m_\rmn{e}}
\end{equation}
and $e$ denotes the elementary charge.

As the regime of strong shocks is dominated by external shocks where
the CR injection is saturated, CRs are dynamically more important in
dilute regions and dynamically less important at the cluster centres
compared to the thermal non-relativistic gas.  As weak shocks are
mainly internal shocks we have to distinguish between their different
appearance: strong internal shocks are most probably accretion shocks
produced by infalling gas from sheets or filaments towards clusters,
or peripheral merger shocks which steepen as they propagate outwards
in the shallow cluster potential, highlighting the importance of CR
injection in the outer cluster regions relative to thermally
dissipated gas at shocks. In contrast, CR injection is dynamically
less important in the case of flow shocks at the cluster centres
or merging shock waves traversing the cluster centre.  From these
considerations we again draw the important conclusion that the ratio
of CR injected energy to dissipated thermal energy at shocks is an
increasing function of decreasing density.  Such a CR distribution
is required within galaxy clusters  to explain the diffuse radio
synchrotron emission of galaxy clusters (so-called radio haloes) within
the hadronic model of secondary electrons. For that, we assume a
stationary CR electron spectrum which balances hadronic injection of
secondaries and synchrotron and inverse Compton cooling processes
\citep{2002BrunettiTaiwan, 2004A&A...413...17P,
2004MNRAS.352...76P}. However, to make more precise statements about
the origin of cluster radio haloes, more work is needed which studies
the effect of the CR dynamics including CR diffusion and other CR
injection processes such as supernovae driven galactic winds.

\subsubsection{Resolution study}
\label{sec:resolution}

To study numerical convergence we perform two additional simulations with a
resolution of $2\times 128^3$, respectively, for our models with and without
reionisation. Fig.~\ref{fig:resolution} shows this resolution study for
non-radiative cosmological simulations with a reionisation epoch at $z=10$.
The lower right-hand panel of Fig.~\ref{fig:resolution} shows the evolution of the
density-weighted temperature with redshift, $\langle(1 + \delta)
T\rangle_V\,(z)$. Shown are different resolutions in our models with and
without reionisation. The two differently resolved simulations ($2\times 256^3$
and $2\times 128^3$) have converged well at redshifts $z \lesssim 4$. In our
reference simulation, the adiabatic decay of the mean temperature is halted at
slightly higher redshift: because of the better mass resolution of this
simulation, nonlinear structures of smaller mass can be resolved earlier while
converting part of their gravitational binding energy into internal energy
through structure formation shock waves.  In the simulation with reionisation,
the temperature increases discontinuously at $z=10$, declines again with the
adiabatic expansion, until shock heating takes over at $z\sim 7-8$ (depending
on the resolution of the simulation).  At $z=0$, all simulations yield a mean
density-weighted temperature of $\simeq 0.3$~keV.  Comparing this
density-weighted energy to the shock-deposited mean energy per particle of
$E_\rmn{diss}\simeq 0.63$~keV, we obtain the mean adiabatic compression factor
of the cosmic plasma, $\{k T / [(\gamma-1) E_\rmn{diss}]\}^{1/(\gamma-1)}
\simeq 0.6$. After the plasma has been shock-heated, relaxation processes in
the course of virialisation let the plasma expand adiabatically on average.
Secondly, mildly non-linear structures characterised by a shallow gravitational
potential are partly effected by the Hubble flow which forces them to
adiabatically expand.

The top left-hand panel shows the differential Mach number distribution
$\dd^2 \eps_\rmn{diss}(a,\M)/( \dd \log a\, \dd \log\M)$ for our
reference simulation with a resolution of $2\times 256^3$ while the
top right-hand panel shows this distribution for the simulation with a
resolution of $2\times 128^3$.  The lower left-hand panel shows both
distributions integrated over the scale factor, $\dd
\eps_\rmn{diss}(\M)/( \dd \log \M)$.  Internal shocks are shown with
dotted lines and external shocks with dashed lines. One immediately
realizes that the question if the first shocks are fully converged
among simulations of different resolution is not well posed because
nonlinear structures of smaller mass can be resolved collapsing
earlier in higher resolution simulations. Accordingly, the
differential Mach number distribution is not well converged at
redshifts $z \gtrsim 6$ while the distribution is well converged for
$z \lesssim 3$.  Since most of the energy is dissipated at late times,
where our differential Mach number distribution is well converged, the
integrated distribution $\dd \eps_\rmn{diss}(\M)/( \dd \log \M)$ shows
only marginal differences among the differently resolved
simulations. In particular, our statements about CR injection at
structure formation shocks are robust with respect to resolution
issues.

\section{Summary and conclusions}
\label{sec:summary}

We provide a formalism for identifying and estimating the strength of structure
formation shocks in cosmological SPH simulations on-the-fly, both for
non-relativistic thermal gas as well as for a plasma composed of a mixture of
cosmic rays (CRs) and thermal gas.  In addition, we derive an analytical
solution to the one-dimensional Riemann shock tube problem for the composite
plasma of CRs and thermal gas (Appendix~\ref{sec:Riemann+CRs}).  In the case of
non-relativistic thermal gas, shock-tube simulations within a periodic
three-dimensional box that is longer in $x$-direction than in the other two
dimensions show that our formalism is able to unambiguously detect and
accurately measure the Mach numbers of shocks, while in the case of plasma
composed of cosmic rays (CRs) and thermal gas, the Mach numbers of shocks are
estimated with an accuracy better than 10\%. In both cases, we find a very good
agreement of the averaged simulated hydrodynamical quantities (such as density,
pressure, and velocity) and the analytical solutions.  Using our formalism for
diffusive shock acceleration, we additionally calculate and monitor the
shock-injected CR energy, but without dynamically tracking this CR energy
component; the latter will be studied in forthcoming work.  The good agreement
between the simulated and theoretically expected shock-injected CR energy
demonstrates that our formalism is reliably able to accelerate CRs
instantaneously during the simulation.

Subsequently, we identified and studied structure formation shock
waves using cosmological N-body/hydrodynamical SPH simulations for a
concordance $\Lambda$CDM universe in a periodic cubic box of comoving
size $100\,h^{-1}$Mpc.  We performed simulations with and without a
reionisation epoch at $z=10$ in order to investigate the effects of
reionisation on the Mach number distribution.  Our sets of simulations
follow only non-radiative gas physics where we neglected additional
physical processes, such as radiative cooling, star formation, and
feedback from galaxies and stars including cosmic ray pressure.  Since
we are mainly interested in shock waves situated mostly outside the
cluster core regions, the conclusions drawn in this article should not
be significantly weakened by the exclusion of those radiative
processes. Furthermore, these simplifications align our work with the
mesh-based simulations of \citet{2003ApJ...593..599R} and enable a
direct comparison and verification of our results.  We classify
cosmological shock waves as internal and external shocks, depending on
whether or not the associated pre-shock gas was previously shocked
\citep[cf.][]{2003ApJ...593..599R}. Rather than using a
thermodynamical criterion such as the temperature, we prefer a density
criterion such as the overdensity $\delta$ in order not to confuse the
shock definition once we will follow radiatively cooling gas in
galaxies.  External shocks surround filaments, sheets, and haloes where
the pristine adiabatically cooling gas is shocked for the first time.
Internal shocks on the other hand are located within the regions bound
by external shocks and are created by flow motions accompanying
hierarchical structure formation. Their population includes accretion
shocks produced by infalling material along the filaments into
clusters, merger shocks resulting from infalling haloes, and flow
shocks inside nonlinear structures which are excited by supersonic
motions of subclumps.

The Mach number distribution weighted by the dissipated energy shows
in detail that most of the energy is dissipated in weak shocks which
are situated in the internal regions of groups or clusters while
collapsed cosmological structures are surrounded by external strong
shocks which have a minor impact on the energy balance.  The evolution
of the Mach number distribution shows that the average shock strength
becomes weaker at later times while there is an increasing amount of
energy dissipated at shock waves as cosmic time evolves because the
mean shock speed increases together with the characteristic mass of
haloes forming during cosmic structure formation. For the same reason,
internal shocks play a more important role than external shocks in
dissipating energy associated with structure formation, especially at
small redshift.  The energy input through reionisation processes
influences the Mach number distribution primarily during a period
following the reionisation era and suppresses strong external shock
waves efficiently because of the significant increase of the sound
speed of the inter-galactic medium.

Weighting the Mach number distribution by the injected CR energy shows the
potential dynamical implications of CR populations in galaxy clusters and
haloes: the maximum and the mean value of the Mach number distribution, weighted
by the shock-dissipated energy, is effectively shifted towards higher values of
the Mach number when the distribution is weighted by the injected CR energy. In
other words, the average shock wave responsible for CR energy injection is
stronger compared to the average shock which thermalizes the plasma.  The
fundamental reason for this lies in the theory of diffusive shock acceleration
at collisionless shock waves and can be phenomenologically described by a CR
injection efficiency: while the CR injection is saturated to an almost
equipartition value between injected CR energy and dissipated thermal energy
for strong shocks, in weak shocks most of the dissipated energy is thermalized
and only small parts are used for the acceleration of relativistic particles.
Relative to the thermal non-relativistic energy density, the shock-injected CR
energy density is dynamically more important at the outer dilute cluster
regions and less important at the cluster centres since weak shock waves
predominantly occur in high-density regions. This has the crucial consequence
that the ratio of CR injected energy to dissipated thermal energy is an
increasing function as the density declines.  Such a CR distribution within
galaxy cluster is required to explain the diffuse radio synchrotron emission of
galaxy clusters (so-called radio haloes) within the hadronic model of secondary
electrons. In order to draw thorough conclusions about the origin of cluster
radio haloes, more work is needed which studies the effect of the CR dynamics
comprising of CR injection and cooling processes as well as CR diffusion
mechanisms.

We note that our new formalism for shock-detection in SPH simulations should
have a range of interesting applications in simulations of galaxy formation.
For example, when combined with radiative dissipation and star formation, our
method can be used to study CR injection by supernova shocks, or to construct
models for shock-induced star formation in the interstellar medium
\citep[e.g.][]{2004MNRAS.350..798B}. It should also be useful to improve the
accuracy of predictions for the production of $\gamma$-rays by intergalactic
shocks \citep[e.g.][]{2003ApJ...585..128K}.

\section*{Acknowledgments}
The authors would like to thank Ewald M{\"u}ller, Bj\"orn Malte Sch\"afer, and
our referee Adrian Jenkins for carefully reading the manuscript and their
constructive remarks.

\bibliography{bibtex/gadget}
\bibliographystyle{mn2e}

\clearpage
\onecolumn
\appendix

\section{Riemann shock-tube problem}
\label{sec:Riemann}

The Riemann shock-tube calculation of \citet{1978JCoPh..27....1S} has
become a generally accepted test of numerical hydrodynamical codes.
As a baseline for later extension, we present in the section the
quasi-analytical solution for the Riemann problem in the standard case
of a polytropic gas. Then, in Appendix~\ref{sec:Riemann+CRs} we derive
the quasi-analytic solution in the case of a gas composed of CRs and
thermal gas, where the effective adiabatic index depends on the
different equations of state and changes across the shock-tube.

In the following, we summarise the key considerations which lead to
the solution of the Riemann problem, for completeness \citep[see
e.g.][ for a compact representation]{1948sfsw.book.....C,
1997rsnm.book.....T, 1991ApJ...377..559R}.  For the initial state, we
assume a state with higher pressure in the left-half space without
loss of generality. At any time $t>0$, this leads to the development
of five regions of gas with different hydrodynamical states which are
numbered in ascending order from the right-hand side. These regions are
separated by the head and the tail of the leftwards propagating
rarefaction wave, and the rightwards propagating contact discontinuity and
the shock wave.  Mass, momentum and energy conservation laws are
represented by the generalised Rankine-Hugoniot conditions for a given
coordinate system:
\begin{eqnarray}
  \label{eq:generalRH}
  \vel_\rmn{d}[\rho] &=& [\rho \vel], \nonumber\\
  \vel_\rmn{d}[\rho \vel] &=& [\rho \vel^2 + P], \\
  \vel_\rmn{d}\left[\rho \frac{\vel^2}{2} + \eps\right] &=& 
  \left[\left(\rho \frac{\vel^2}{2} + \eps + P\right)\vel\right]. \nonumber
\end{eqnarray}
Here $\vel_\rmn{d}$ denotes the speed of the discontinuity under consideration
with respect to our coordinate system and we introduced the abbreviation $[F] =
F_i - F_j$ for the jump of some quantity $F$ across the discontinuity.  Within
the leftwards propagating rarefaction wave, the generalised Riemann invariants
yield an isentropic change of state, $\dd s = 0$, and conserve the quantity
$\Gamma^+$:
\begin{equation}
  \label{eq:Riemann}
  \Gamma^+ = \vel + \int_0^\rho \frac{c(\rho')}{\rho'}\dd \rho'
   = \vel + \frac{2\, c(\rho)}{\gamma - 1} = \rmn{const}.
\end{equation}
For the last step, we assumed a polytropic equation of state $P = A
\rho^\gamma$. Appropriately combining these equations, the solution can be
expressed as follows:
\begin{equation}
  \label{eq:rho}
  \rho(x, t) = 
  \left\{ \begin{array}{ll}
      \rho_5, & x \le -c_5 t, \\
      \rho_5 \left[-\mu^2 \frac{\dps x}{\dps c_5 t} + (1 - \mu^2)\right]^{2/(\gamma-1)}, 
      & -c_5 t < x \le -\vel_\rmn{t} t, \\
      \rho_3, & -\vel_\rmn{t} t < x \le \vel_2 t, \\
      \rho_2, & \vel_2 t < x \le \vel_\rmn{s} t, \\
      \rho_1, & x > \vel_\rmn{s} t, \\
    \end{array} \right.
\end{equation}
\begin{equation}
  \label{eq:P}
  P(x, t) = 
  \left\{ \begin{array}{ll}
      P_5, & x \le -c_5 t, \\
      P_5 \left[-\mu^2 \frac{\dps x}{\dps c_5 t} + (1 - \mu^2)\right]^{2\gamma/(\gamma-1)}, 
      & -c_5 t < x \le -\vel_\rmn{t} t, \\
      P_2 = P_3, & -\vel_\rmn{t} t < x \le \vel_\rmn{s} t, \\
      P_1, & x > \vel_\rmn{s} t, \\
    \end{array} \right.
\end{equation}
\begin{equation}
  \label{eq:vel}
  \vel(x, t) = 
  \left\{ \begin{array}{ll}
      0, & x \le -c_5 t, \\
      (1 - \mu^2)\left( \frac{\dps x}{\dps t} + c_5\right),
      & -c_5 t < x \le -\vel_\rmn{t} t, \\
      \vel_2 = \vel_3, & -\vel_\rmn{t} t < x \le \vel_\rmn{s} t, \\
      0, & x > \vel_\rmn{s} t. \\
    \end{array} \right.
\end{equation}
Here $\mu^2 = (\gamma - 1) / (\gamma + 1)$, $c_1 = \sqrt{\gamma P_1 / \rho_1}$,
and $c_5 = \sqrt{\gamma P_5 / \rho_5}$ are the speeds of sound, $\vel_\rmn{t}$
is the speed of propagation of the rarefaction wave's tail, and $\vel_\rmn{s}$
is the shock speed. The post-shock pressure is obtained by solving (numerically)
the non-linear equation, which is derived from the Rankine-Hugoniot conditions
over the shock while ensuring the conservation of the two Riemann invariants of
equation~(\ref{eq:Riemann}):
\begin{equation}
  \label{eq:P3}
  \left(\frac{P_2}{P_1} - 1\right)
  \sqrt{\frac{1 - \mu^2}{\gamma\, (P_2 / P_1 + \mu^2)}}
  - \frac{2}{(\gamma - 1)}\frac{c_5}{c_1} 
  \left[1 - \left(\frac{P_2}{P_5}\right)^{(\gamma-1)/(2\gamma)}\right] = 0.
\end{equation}
The density on the left-hand side of the contact discontinuity is $\rho_3 =
\rho_5 (P_2 / P_5)^{1 / \gamma}$, since the gas is adiabatically connected to
the left-hand side. The post-shock density $\rho_2$ is also derived from the
Rankine-Hugoniot conditions,
\begin{equation}
  \label{eq:rho4}
  \rho_2 = \rho_1\left(\frac{P_2 + \mu^2 P_1}{P_1 + \mu^2 P_2}\right).
\end{equation}
The post-shock gas velocity $\vel_2$ is obtained from the rarefaction wave
equation, $x / t = \vel - c$, and usage of the Riemann invariant $\Gamma^+$:
\begin{equation}
  \label{eq:v3}
  \vel_2 = \vel_3 = \frac{2 c_5}{(\gamma - 1)}
  \left[1 - \left(\frac{P_2}{P_5}\right)^{(\gamma - 1) / (2\gamma)}\right],
\end{equation}
and from equation~(\ref{eq:vel}) we derive the speed of propagation of the
rarefaction wave's tail $\vel_\rmn{t} = c_5 - \vel_2 / (1 - \mu^2)$.  Finally,
mass conservation across the shock yields
\begin{equation}
  \label{eq:vs}
  \vel_\rmn{s} = \frac{\vel_\rmn{2}}{1 - \rho_1 / \rho_2}.
\end{equation}

\section{Riemann shock-tube problem for a composite of cosmic rays and thermal gas}
\label{sec:Riemann+CRs}

\subsection{Derivation}
In contrast to the previous case, the composite of CRs and thermal gas does not
obey a polytropic equation of state.  In this section, we present an analytical
derivation of the Riemann shock-tube problem for the composite of polytropic
gas and a component that is adiabatically compressed at the shock such as
relativistic gas or a homogeneous magnetic field which is parallel to the shock
front. For the analytical derivation, we adopt the following two
approximations: (i) we assume the CR adiabatic index (equation~\ref{eq:gammaCR})
to be constant over the shock-tube, and (ii) we neglect CR diffusion.  The
first assumption is justified as long as the CR pressure is not dominated by
trans-relativistic CRs of low energy while the second assumption is a strong
simplification with respect to simulating realistic shocks including CRs
\citep{2005ApJ...620...44K}. However, including CR diffusion complicates the
problem significantly such that it is not any more analytically tractable.

For the initial state, we again assume a state with higher pressure in the
left-half space. At any time $t>0$, five regions of gas with different
hydrodynamical states coexist, and are numbered in ascending order from the
right-hand side. We use the notation $P_i = P_{\CR,i} + P_{\th,i}$ and $\eps_i
= \eps_{\CR,i} + \eps_{\th,i}$ for the composite quantities in region $i$. The
full solution of the initial value problem consists of determining 12 unknown
quantities in the regions (2) and (3): $\rho_2$, $\vel_2$, $P_{\CR2}$,
$P_{\th2}$, $\eps_{\CR2}$, $\eps_{\th2}$, and $\rho_3$, $\vel_3$, $P_{\CR3}$,
$P_{\th3}$, $\eps_{\CR3}$, $\eps_{\th3}$. The thermal gas obeys a polytropic
equation of state, i.e.~$\eps_{\th,i} = P_{\th,i} / (\gamma_\th - 1)$ for $i
\in \{2,3\}$ and the regions (2) and (3) are separated by a contact
discontinuity, implying vanishing mass flux through it and thus, $\vel_2 =
\vel_3$ and $P_2 = P_3$.  This reduces the dimensionality of our problem to 8
unknowns.  In our approximation, the CRs are adiabatically expanded over the
rarefaction wave and adiabatically compressed at the shock while obeying a
polytropic equation of state:
\begin{equation}
  \label{eq:CRadiabatic}
  \begin{array}{lcl lcl}
    P_{\CR3} & = & 
    P_{\CR5} \left(\frac{\dps\rho_3}{\dps\rho_5}\right)^{\gamma_\CR},
    \quad &
    \eps_{\CR3} & = & 
    \eps_{\CR5} \left(\frac{\dps\rho_3}{\dps\rho_5}\right)^{\gamma_\CR},
    \\ \rule{0cm}{0.6cm}
    P_{\CR2} & = & 
    P_{\CR1} \left(\frac{\dps\rho_2}{\dps\rho_1}\right)^{\gamma_\CR},
    \quad &
    \eps_{\CR2} & = & 
    \eps_{\CR1} \left(\frac{\dps\rho_2}{\dps\rho_1}\right)^{\gamma_\CR},
  \end{array}
\end{equation}
which further reduces the dimensionality by 4 unknowns. Moreover, the thermal
gas is also adiabatically expanded over the rarefaction wave yielding
$P_{\th3} = P_{\th5} (\rho_3/\rho_5)^{\gamma_\th}$. Hence, we need 3 more
linearly independent equations for the solution: 2 are obtained by considering
the Rankine-Hugoniot conditions (equation~\ref{eq:generalRH}) in a stationary
system of reference with $\vel_\rmn{d} = \vel_\rmn{s}$. The last equation is
given by the Riemann invariant $\Gamma^+$, where the effective speed of sound
is given by $c = \sqrt{\gamma_{\eff} P / \rho}$:
\begin{equation}
  \label{eq:RiemannCR}
  \Gamma^+ = \vel + \int_0^\rho \frac{c(\rho')}{\rho'}\dd \rho'
  = \vel + I(\rho) = \rmn{const.}
  \quad\mbox{with}\quad
  I(\rho) = \int_0^\rho 
  \sqrt{\A_\CR x^{\gamma_\CR - 3} + \A_\th x^{\gamma_\th - 3}} \dd x.
\end{equation}
Here, we use the abbreviations $\A_i = \gamma_i A_i$ where $i \in
\{\rmn{th},\rmn{CR}\}$ and $A_{i}^{} = P_{i}^{}\,\rho^{-\gamma_i}$ denotes the
invariant adiabatic function over the rarefaction wave. Introducing the
difference of the adiabatic indices of the two populations, $\Delta\gamma =
\gamma_\th - \gamma_\CR$, the solution to the integral $I(\rho)$ is given by
\begin{equation}
  \label{eq:RIsolution}
  I(\rho) = \frac{\sqrt{\A_\CR}}{\Delta\gamma}
  \left(\frac{\A_\CR}{\A_\th}\right)^{(\gamma_\CR - 1)/(2 \Delta\gamma)}
  \B_{x(\rho)} \left(\frac{\gamma_\CR - 1}{2\Delta\gamma},
             \frac{1 - \gamma_\th}{2\Delta\gamma}\right)
  \quad\mbox{with}\quad
  x(\rho) = \frac{\A_\th\, \rho^{\gamma_\th}}
  {\A_\CR\, \rho^{\gamma_\CR} + \A_\th\, \rho^{\gamma_\th}}.
\end{equation}
Although the second argument of the incomplete Beta-function is always
negative, $I(\rho)$ is well defined as long as we consider a non-zero CR
pressure which is characterised by $\A_\CR > 0$, and $\gamma_\CR$ sufficiently
far from $\gamma_\th$, i.e.~ $\Delta\gamma > 0$. For $\A_\CR = 0$, the integral
can be solved in closed form, yielding $I(\rho) = 2 c(\rho) / (\gamma_\th-1)$.

\subsection{Solution of the Riemann problem}

The densities leftwards and rightwards of the contact discontinuity, $\rho_3$
and $\rho_2$, are obtained by solving (numerically) the following non-linear
system of equations. It is derived from matching the possible post-shock states
(pressure and density) with the possible post-rarefaction wave states while
simultaneously ensuring the conservation laws over the rarefaction wave and
the shock:
\begin{equation}
  \label{eq:xs,xr}
  \begin{array}{lclcl}
    f_1(x_\rmn{s},x_\rmn{r}) & \equiv &
    [P_2(x_\rmn{r}) - P_1]\, (x_\rmn{s} - 1) - 
    \rho_1 x_\rmn{s} \left[ I(\rho_5) - I(x_\rmn{r}\rho_5) \right]^2 & = & 0,
    \\ \rule{0cm}{0.4cm}
    f_2(x_\rmn{s},x_\rmn{r}) &\equiv &
    [P_2(x_\rmn{r}) + P_1]\, (x_\rmn{s} - 1) + 
    2 [x_\rmn{s} \eps_1 - \eps_2(x_\rmn{s},x_\rmn{r})] & = & 0.
  \end{array}
\end{equation}
Here we introduced the shock compression ratio $x_\rmn{s} \equiv \rho_2 /
\rho_1$ and the rarefaction ratio $x_\rmn{r} \equiv \rho_3 / \rho_5$.
Furthermore, the implicit dependences on $x_\rmn{s}$ and $x_\rmn{r}$ can
explicitly be expressed as follows,
\begin{eqnarray}
  \label{eq:abbr:P3}
  P_2(x_\rmn{r}) &=& P_3(x_\rmn{r}) = P_{\CR5} x_\rmn{r}^{\gamma_\CR} + 
                     P_{\th5} x_\rmn{r}^{\gamma_\th}, \\
  P_{\CR2}  (x_\rmn{s}) &=& P_{\rmn{\CR1}} x_\rmn{s}^{\gamma_\CR}, \\ 
  \eps_2(x_\rmn{s},x_\rmn{r}) &=& \eps_{\CR1} x_\rmn{s}^{\gamma_\CR} + 
  \frac{1}{\gamma_\th-1}[P_2(x_\rmn{r}) - P_{\CR2}(x_\rmn{s})].
\end{eqnarray}
The roots of the non-linear system of equations (equation~\ref{eq:xs,xr})
immediately yield the post-shock pressure of the fluid via
equation~(\ref{eq:abbr:P3}). The post-shock velocity $\vel_2 = \vel_3$ and the
shock speed $\vel_\rmn{s}$ are then obtained from the Rankine-Hugoniot
relations,
\begin{eqnarray}
  \label{eq:v3CR}
  \vel_2 &=& 
  \sqrt{[P_2(x_\rmn{r}) - P_1]\, \frac{\rho_2 - \rho_1}{\rho_2 \rho_1}}, \\
  \vel_\rmn{s} &=& \frac{\rho_2 \vel_2}{\rho_2 - \rho_1}.
\end{eqnarray}
\newpage

Using the previous results, we can construct the solution to the generalised
Riemann problem for CRs and thermal gas as follows:
\begin{equation}
  \label{eq:rhoCR}
  \rho(x, t) = 
  \left\{ \begin{array}{ll}
      \rho_5, & x \le -c_5 t, \\
      \rho(x,t), & -c_5 t < x \le -\vel_\rmn{t} t, \\
      \rho_3, & -\vel_\rmn{t} t < x \le \vel_2 t, \\
      \rho_2, & \vel_2 t < x \le \vel_\rmn{s} t, \\
      \rho_1, & x > \vel_\rmn{s} t, \\
    \end{array} \right.
\end{equation}
\begin{equation}
  \label{eq:PCR}
  P(x, t) = 
  \left\{ \begin{array}{ll}
      P_5, & x \le -c_5 t, \\
      A_\CR\, \rho(x,t)^{\gamma_\CR} + A_\th\, \rho(x,t)^{\gamma_\th}, 
      & -c_5 t < x \le -\vel_\rmn{t} t, \\
      P_2 = P_3, & -\vel_\rmn{t} t < x \le \vel_\rmn{s} t, \\
      P_1, & x > \vel_\rmn{s} t, \\
    \end{array} \right.
\end{equation}
\begin{equation}
  \label{eq:velCR}
  \vel(x, t) = 
  \left\{ \begin{array}{ll}
      0, & x \le -c_5 t, \\
      \frac{\dps x}{\dps t} + 
      \sqrt{\A_\CR\, \rho(x,t)^{\gamma_\CR-1} + \A_\th\, \rho(x,t)^{\gamma_\th-1}},
      & -c_5 t < x \le -\vel_\rmn{t} t, \\
      \vel_2 = \vel_3, & -\vel_\rmn{t} t < x \le \vel_\rmn{s} t, \\
      0, & x > \vel_\rmn{s} t. \\
    \end{array} \right.
\end{equation}
Here $c_5 = \sqrt{\gamma_{\eff5} P_5 / \rho_5}$ is the effective speed of sound,
$\vel_\rmn{t}$ is the speed of propagation of the rarefaction wave's tail, and
$\vel_\rmn{s}$ is the shock speed.  Matching the  rarefaction wave equation to
the density of the post-contact discontinuity yields $\vel_\rmn{t}$:
\begin{equation}
  \label{eq:vtCR}
  \vel_\rmn{t} = I(\rho_3) - I(\rho_5) +
  \sqrt{\A_\CR\, \rho_3^{\gamma_\CR-1} + \A_\th\, \rho_3^{\gamma_\th-1}}.
\end{equation}
The density within the rarefaction regime is obtained by solving (numerically)
the non-linear equation for a given $(x,t)$, which is derived from the
rarefaction wave equation,
\begin{equation}
  \label{eq:rfCR}
  I[\rho(x,t)] - I(\rho_5) + \frac{\dps x}{\dps t} + 
  \sqrt{\A_\CR\, \rho(x,t)^{\gamma_\CR-1} + \A_\th\, \rho(x,t)^{\gamma_\th-1}}
  = 0.
\end{equation}

\bsp

\label{lastpage}

\end{document}